%% file: paper.tex
\documentclass[twocolumn,conference,10pt,comsocconf]{IEEEtran}
\usepackage{algorithm,subfigure}
\usepackage{algorithmic}
\usepackage{algorithm}
\usepackage{amsmath,color}
\usepackage{amsthm}
\usepackage{graphicx}
\usepackage{amssymb}
\usepackage{url}
\usepackage{color}
\usepackage{multirow}

\newcommand {\tr}{Appendix}

\newtheorem{theorem}{Theorem}

\def\done{\hspace*{\fill} \rule{1.8mm}{2.5mm} \\ }

\begin{document}

\title{Stochastic Modeling of Hybrid Cache Systems}

\author{\IEEEauthorblockN{Gaoying Ju$^1$, Yongkun Li$^{1,2}$, Yinlong Xu$^{1,3}$, Jiqiang Chen$^1$, John C. S. Lui$^4$}
$^{1}$School of Computer Science and Technology, University of Science and
Technology of China\\
$^{2}$Collaborative Innovation Center of High Performance Computing, National University of Defense Technology\\
$^{3}$AnHui Province Key Laboratory of High Performance Computing\\
$^{4}$Department of Computer Science and Engineering, The Chinese University of Hong Kong\\
\{jgy93317, cjqld\}@mail.ustc.edu.cn, \{ykli, ylxu\}@ustc.edu.cn,
cslui@cse.cuhk.edu.hk}


\maketitle
\input{abstract}
\input{introduction}
\input{designspace}
\input{systemmodel}
\input{meanfield}
\input{validation}
\input{numerical}
\input{relatedwork}
\input{conclusion}
\input{ack}
\begin{small}
\bibliographystyle{abbrv}
\bibliography{paper}
\end{small}
\input{app}
\end{document}

%% file: abstract.tex
\begin{abstract}

In recent years, there is an increasing demand of big memory systems so to
perform large scale data analytics.  Since DRAM memories are expensive,
some researchers are suggesting to use other memory systems such as
non-volatile memory (NVM) technology to build large-memory computing systems.
However, whether the NVM technology can be a viable alternative (either
economically and technically) to DRAM remains an open question. To answer
this question, it is important to consider how to design a memory system
from a ``system perspective'', that is, incorporating different
performance characteristics and price ratios from hybrid memory devices.

This paper presents an analytical model of a ``{\em hybrid page cache
system}'' so to understand the diverse design space and performance impact
of a hybrid cache system.  We consider (1) various architectural choices,
(2) design strategies, and (3) configuration of different memory devices.
Using this model, we provide guidelines on how to design hybrid page cache
to reach a good trade-off between high system throughput (in I/O per sec
or IOPS) and fast cache reactivity which is defined by the time to fill
the cache. We also show how one can configure the DRAM capacity and NVM
capacity under a fixed budget. We pick PCM as an example for NVM and conduct numerical analysis.
Our analysis indicates that incorporating PCM in a page cache system significantly improves the system performance,
and it also shows  larger benefit  to allocate more PCM in page cache in
some cases. Besides, for the common setting of performance-price ratio of
PCM, ``flat architecture'' offers as a better choice, but ``layered
architecture'' outperforms if PCM write performance can be significantly
improved in the future.


\end{abstract}


%% file: introduction.tex
\section{Introduction}
\label{sec:introduction}

In modern computer systems, there is a common consensus that secondary
storage devices such as hard disk drives (HDDs) are orders of magnitude
slower than memory devices like DRAM. Even though flash-based storage
devices like solid-state drives (SSDs), which are much faster than HDDs,
have been quickly developed and widely used in recent years, they cannot
replace DRAM since SSDs have lower I/O throughput than DRAM (i.e.,  at least
an order of magnitude lower). Due to the large performance gap between
memory and secondary storage, I/O access poses as a major bottleneck for
computer system performance. To address this issue, one commonly used
technique is to allow some memory as page cache, which exploits workload
locality by buffering the recently accessed data in fast-speed memory for a
short time before flushing to the slow-speed storage devices. Using page
caches, one can mitigate the performance mismatch between memory and
storage.

%
%

Traditional page cache usually uses DRAM due to its high throughput (in
terms of IOPS), e.g., \cite{bovet2005understanding, Kgil2006FlashCache,
lee2014eliminating}. However, solely relying on DRAM has at least three
limitations.
First, the development of DRAM technology has already reached its limit,
e.g., DRAM scaling is more difficult as charge storage and sensing
mechanisms will become less reliable when scaled to thinner manufacturing
processes \cite{raoux2008phase}. Second, the price of DRAM is still much
higher than that of HDDs or SSDs, and it also consumes much more energy due
to its refresh operations. So  DRAM-based main memory consumes a significant
portion of the total system cost and energy with its increasing size
\cite{kgil2008improving}. Finally, DRAM is a volatile  device and data in
DRAM will disappear if there is any power failure. Hence, keeping a lot of
data in DRAM implies  lowering the system reliability.


Non-volatile memory (NVM) technologies (e.g. PCM, STT-MRAM, ReRAM)
offer an alternative to DRAM due to their byte-addressable feature (which is similar to DRAM) and higher
throughput than flash memory. In particular, NVM is commonly accepted as a
new tier in the storage hierarchy ``{\em between}'' DRAM and SSDs, and it also poses a
design trade-off when we use it as page cache. On the one hand, it is much
faster than flash-based SSDs  but still slower than DRAM, so replacing DRAM
with NVM in page cache may degrade the system performance. On the other
hand, the price and single-device capacity of NVM are also considered to lie
between DRAM and SSDs, so one can have more NVM storage capacity than DRAM
given a fixed budget. Furthermore, due to the non-volatile property of NVM, even
keeping a large amount of data in NVM does not reduce the system
reliability. Thus, it is possible to have a large page cache with NVM, which
 increases the cache hit ratio and as a result
 improves the overall system performance. Therefore, it remains an open
 question  whether it is more efficient to
consider a hybrid cache system with both DRAM and NVM,  and how to fully
utilize the benefits of NVM in page cache design. {\em This motivates us to
develop a mathematical model to comprehensively study the impact of
architecture design and system configurations on page cache performance, and
explore the full design space when both DRAM and NVM are available.}

%

However, analyzing a hybrid cache system is challenging.  First,
including NVM in page cache clearly introduces system heterogeneity, and so
it offers more choices for system design and severely increases the
analysis complexity. For example, when both DRAM and NVM are used, should we
consider a ``{\em flat architecture}'' which places DRAM and NVM in the same
level  and accesses them in parallel, or consider a ``{\em layered
architecture}'' which uses DRAM as a cache for NVM? Another question is how to
allocate the capacity of each device under a fixed budget so as to maximize
the system performance.
Second, since access to DRAM and NVM have different latencies, it is not
accurate to analyze the system performance by deriving only the hit ratio as
in traditional cache analysis. In fact, one needs to explicitly take the
difference of latency into account in the analysis. We emphasize that
measurement studies with simulator/prototype are also feasible methods, but
they may suffer from the efficiency problem due to the wide
choices in system design. While analytical modeling is easy to be
parameterized and generally needs less running time.



Motivated by the list-based model developed
by Gast et al. in \cite{transient15},  in this paper, we extend the
 model to analyze hybrid cache systems under both the flat and layered
architectures. We also take into consideration the device heterogeneity by
defining a latency-based model to characterize the cache performance so as
to explore the full design space and the optimal architecture design. To the
best of our knowledge, this is the first work which uses mathematical models
to analyze hybrid cache systems with DRAM and NVM.

The main contributions of this paper are as follows.

\begin{itemize}

 \item We extend the list-based model in \cite{transient15} to
     characterize the dynamics of cache content distribution in hybrid
     cache systems under both flat and layered architectures, and derive
     the steady-state solution by using  a mean-filed approximation. We
     make each device operate in a fine granularity by dividing it into
     multiple lists with a layered structure so as to explore the optimal
     system performance and full design space.

 \item We propose a latency-based metric to quantify the hybrid cache
     performance. To support the latency model, we conduct measurements in
     the Linux kernel level to obtain the average request delay at the
     granularity of nanoseconds. With this latency model, we are able to
     take the heterogeneity of different devices into account so as to
     study the impact of different design choices on hybrid cache
     performance with higher accuracy.

    \item We validate our analysis with simulations  by modifying the
        DRAMSim2 simulator \cite{dramsim2}. We further study the impact of
        different architectures (flat or layered) and different system
        settings, such as the number of lists in each cache device, the
        performance-price ratio of NVM, as well as the capacity allocation
        of each cache device, on the hybrid cache performance via
        numerical analysis.

    \item Our analysis results show that incorporating PCM in hybrid cache
        design significantly improves the system performance over
        traditional DRAM-only cache under the common setting of
        performance-price ratio. Furthermore, the hybrid cache design
        needs to be adjusted accordingly when the ratio varies. In
        particular, the number of lists in each cache device should be
        configured carefully to achieve a good trade-off between the cache
        performance and cache reactivity. Besides, under the common
        setting of performance-price ratio of PCM, flat architecture
        offers  a better choice, but layered architecture outperforms if PCM write
        performance gets significantly improved.

\end{itemize}


The rest of this paper proceeds as follows. 
In \S\ref{sec:design_choices}, we introduce the architecture design and
system configurations of hybrid page cache, and formulate multiple design
issues to motivate our study. We present the Markov model for characterizing
the cache content distribution in \S\ref{sec:systemmodel}, and derive the
mean-field approximation in \S\ref{sec:meanfield}. We validate our analysis
by using DRAMSim2 simulator in \S\ref{sec:validation}, and show the analysis
results and insights via numerical analysis in \S\ref{sec:numerical}.
Finally, we review related work in \S\ref{sec:relatedwork}, and conclude the
paper in \S\ref{sec:conclusion}.



%% file: designspace.tex
\section{Design Choices and Issues of Hybrid Cache} \label{sec:design_choices}


In this section, we first introduce the system architecture and design
choices of hybrid cache systems that we analyze in this paper. In
particular, we consider two types of system architectures: flat architecture
and layered architecture (see \S\ref{subsec:architectures}), and study a
fine-grained list-based cache replacement algorithm (see
\S\ref{subsec:replacement_alg}). After that, we formulate several design
issues to motivate our study (see \S\ref{subsec:issues}).

\subsection{System Architecture}\label{subsec:architectures}

We focus on hybrid cache design which composes of both DRAM and NVM. For
ease of presentation, we call DRAM and NVM used in a cache {\em D-Cache} and
{\em N-Cache}, respectively, and assume that we have $m_D$ DRAM pages and
$m_N$ NVM pages with the same page size, say 4KB, in the system. That is,
the capacity of D-Cache is $m_D$, and that of N-Cache is $m_N$. We also
denote $m$ as the total capacity of the hybrid cache, i.e., $m=m_D+m_N$. We
denote the overall system cost as $C=m_D*c_D+m_N*c_N$, where $c_D$ and $c_N$
denote the price/cost of each page of DRAM and NVM,
respectively. 



To organize D-Cache and N-Cache, we further divide each of them into
multiple lists, each of which contains a certain number of pages, and denote
the number of lists in D-Cache and N-Cache as $h_D$ and $h_N$, respectively.
We label the  lists of N-Cache as $l_{1}, \cdots, l_{h_N}$, and label the
lists of D-Cache as $l_{h_N+1}, \cdots, l_{h}$, where $h=h_N+h_D$ denotes
the total number of lists in the whole system.
For list $l_i$, we define its capacity as $m_i$, so we have
and we have $\mathbf{m}=(m_1,...,m_h)$, with $\sum_{i=1}^{h}m_i=m$, which describes the whole cache system.

\begin{figure}[!ht]
\centering
\subfigure[Flat Architecture]
{\includegraphics[width=0.53\linewidth]{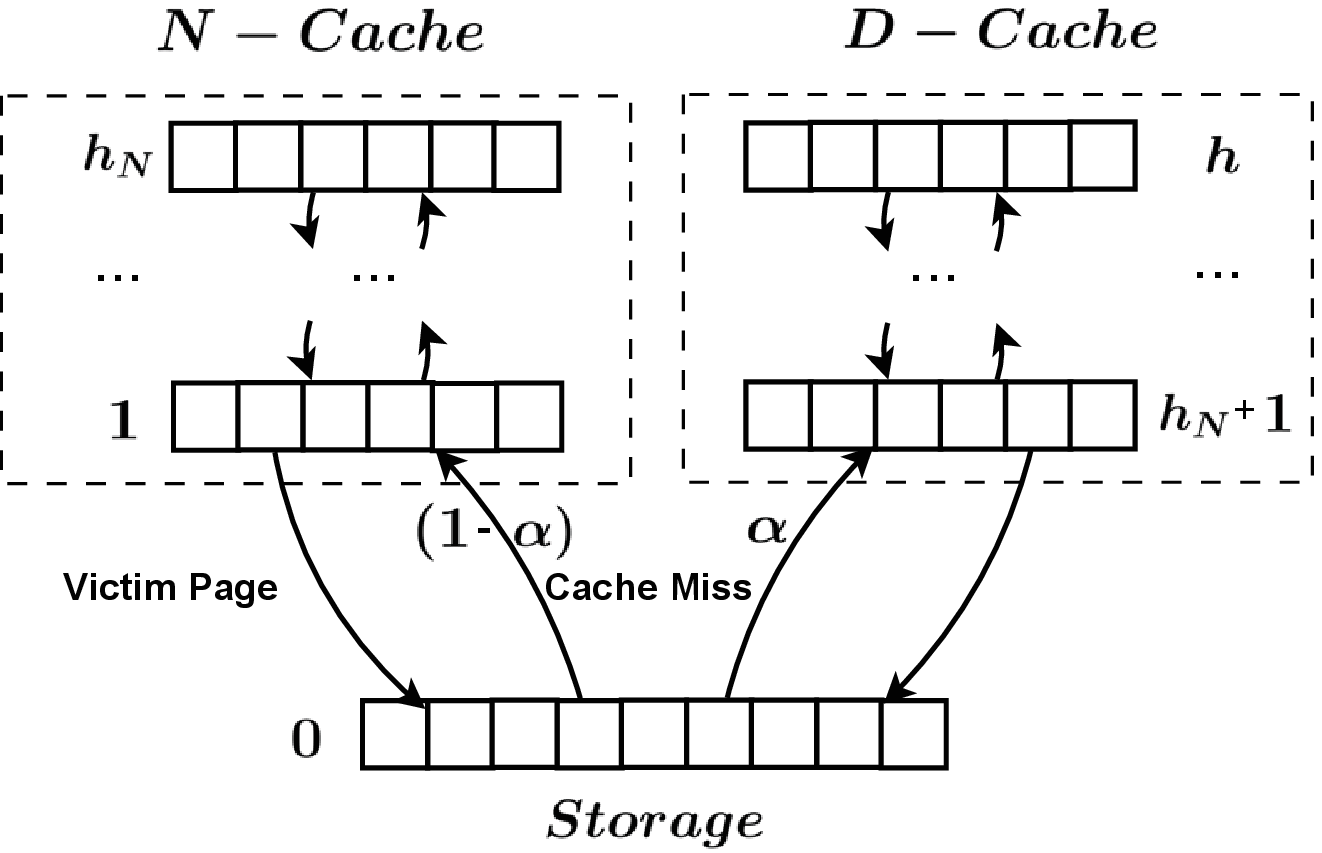}}
\subfigure[Layered Architecture]
{\includegraphics[width=0.43\linewidth]{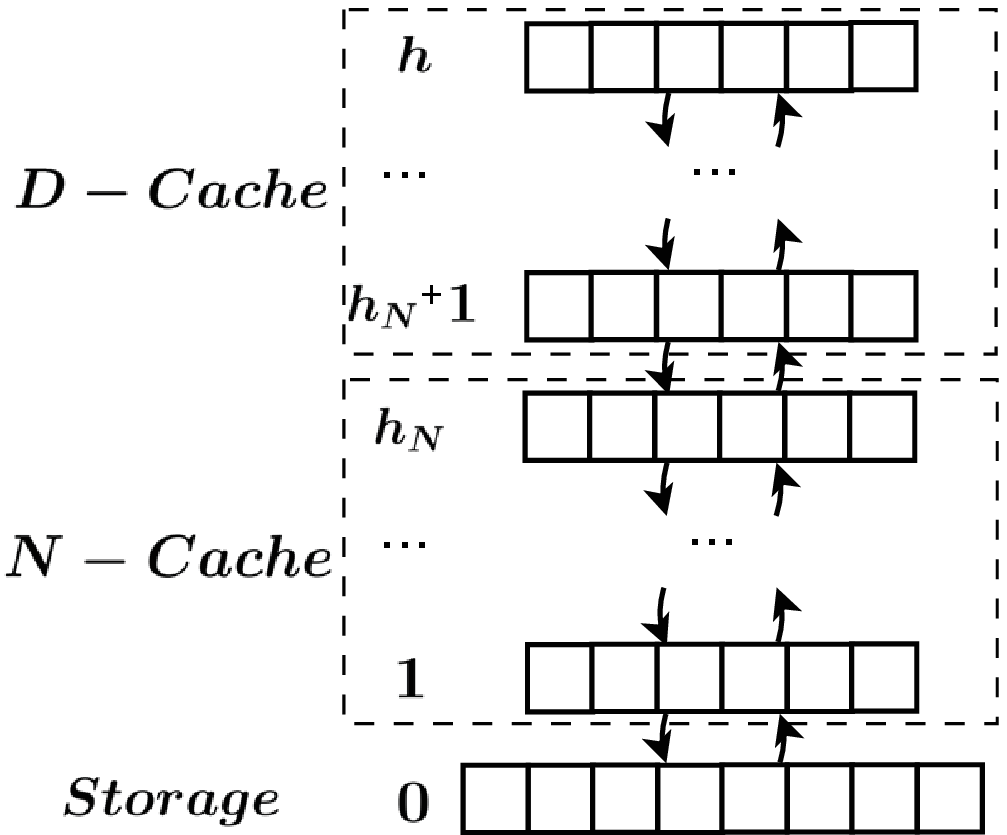}}
\vskip -5pt
\caption{Architecture of hybrid cache.}
\vskip -15pt
\label{fig:arctecture}
\end{figure}

We denote the secondary storage layer as list $l_{0}$. Without loss of
generality, we  call list $l_i$ the $i$-th list, i.e., $l_i=i$.
Figure~\ref{fig:arctecture} shows an example of the list-based organization
of D-Cache and N-Cache under different architectures.

To design a hybrid cache with both D-Cache and N-Cache, we consider two
architectures: flat architecture and layered architecture, which are
described as follows.
\begin{itemize}
  \item \emph{Flat architecture:} In this design, both D-Cache and N-Cache
      are placed in the same level and accessed in parallel as shown in
      Figure~\ref{fig:arctecture}(a). In particular, for a new data page
      which has not been cached before, it is either cached in D-Cache
      with probability $\alpha$ or in N-Cache with probability $1-\alpha$.
      Note that $\alpha$ is a tunable parameter, and increasing it implies
      that D-Cache is more preferred to be used. In the flat architecture,
       pages are never migrated between the two types of caches.

      Note that both D-Cache and N-Cache contain multiple lists. To
    exploit workload locality, we let pages be first buffered in the list with
    the smallest label in the corresponding cache, and then upgrade to the
    larger-numbered lists when they become hot (e.g., when cache hit
      happens). That is, lists in the same cache device are organized in a
      layered structure.
  \item \emph{Layered architecture:} In this design, we use D-Cache as a
      caching layer for N-Cache as shown in
      Figure~\ref{fig:arctecture}(b). Particularly, new data page is
      directly buffered in N-Cache first, and when page in the list of
      the largest label in N-Cache is accessed, it is upgraded to
      D-Cache. Similarly, we also organize lists in both D-Cache and
      N-Cache in a layered structure. Note that
     data  migration  between D-Cache and N-Cache happens here,
     and usually, data in D-Cache is considered to be hotter than data in
     N-Cache.
\end{itemize}
%

\subsection{Cache Replacement Algorithm}\label{subsec:replacement_alg}

For cache replacement, we follow the
list-based algorithm introduced in \cite{transient15}, and extend it to
hybrid cache with different architectures.
Roughly speaking, a new
data page enters into a cache through the first list and  moves to the upper
list by exchanging with a randomly selected data page whenever a cache hit
occurs. Specifically, when a data page $k$ is requested at time $t$, one of
the three events below happens:

\begin{itemize}
  \item {\emph{Cache miss:}} Page $k$ is not in D-Cache nor N-Cache. In
      this case, page $k$ enters into the first list in D-Cache (i.e.,
      list $l_{h_N+1}$) with probability $\alpha$ or into the first list
      in N-Cache (i.e., list $l_{1}$)  with probability $(1-\alpha)$ under
      the flat architecture. For the layered architecture, page $k$ enters
      into the first list of N-Cache (i.e., list $l_{1}$). For both
      architectures, the position in the list for writing page $k$ is
      chosen uniformly at random. Meanwhile, the victim page in the position
      moves back to list 0.

  \item {\emph{Cache hit in list $l_i$ where $l_i \neq l_{h_{N}}$ and
      $l_i\neq l_h$}:} In this case, page $k$ moves to a randomly selected
      position $v$ of list $l_{i+1}$, meanwhile, the victim page in
      position $v$ of list $l_{i+1}$ takes the former position of page
      $k$.

  \item {\emph{Cache hit in list $l_i$ where $l_i = l_{h_{N}}$ or $l_i =
      l_h$}:}  In this case, page $k$  remains at the same position under
      the flat architecture. However,  for the layered architecture with
      $l_i = l_{h_{N}}$, page $k$ moves to a random  position in list
      $l_{i+1}$ as in the second case.
\end{itemize}

Figure~\ref{fig:arctecture} shows the data flow under flat and layered
{architectures}. Note that data migration happens between lists of the same
type of cache, while the migration between D-Cache and N-Cache happens only
in the case of  layered architecture.

\subsection{Design Issues}\label{subsec:issues}
Note that the overall performance of a hybrid cache system may depend on
various factors, such as system architecture, capacity allocation between
DRAM and NVM, as well as the configuration parameters like the number of
lists in each cache device. Thus, it poses a wide range of design choices
for hybrid cache, which makes it very difficult to explore the full design
space and optimize the cache performance. To understand the impact of hybrid
cache design on system performance, in this work, we aim to address the
following issues by developing mathematical models.

\begin{itemize}

  \item For each architecture (flat or layered), what is the impact of the
      list-based hierarchical design, and how to set the best parameters
      so as to optimize the overall performance, including the {numbers}
      of lists $h_D$  and $h_N$, as well as the preference parameter
      $\alpha$ for the flat architecture?

\item Which architecture should be used when considering both DRAM and NVM
    into a hybrid design?

  \item Under a fixed budget $C$, what is the best capacity allocation of
      each cache type for better performance?

\end{itemize}

%% file: systemmodel.tex
\section{System Model} \label{sec:systemmodel}

 In this section, we first describe the workload model, then  characterize the dynamics of data pages in hybrid cache,
 and finally derive the cache content distribution in steady state. After
 that, we define a latency-based performance metric based on the cache content distribution
 so as to quantify the overall cache performance.

\subsection{Workload Model}\label{subsec:workload}

In this work, we focus on  cache-effective applications like web search and
database query \cite{zheng2011fastscale,Kgil2006FlashCache}, in which memory
and I/O latency are critical to system performance. Thus, caching files in
main memory becomes necessary to provide sufficient throughput for these
applications. To provide high data reliability, we assume to use  the
write-through policy, in which data is also written to the storage tier once
it is buffered in the page cache. With this policy, all data pages in cache
should have a copy in the secondary storage.

In this paper, we focus on the independent reference model
\cite{transient15} in which requests in a workload are independent of each
other. Since cache mainly benefits the read performance, we focus on read
requests only, while we can also extend our model to write requests. Suppose
that we have $n$ total data pages in the system. In each time slot, one read
request arrives, and  it accesses data pages according to a particular
distribution where page $k$ ($k$ = 1, 2, ..., $n$) is accessed with
probability $p_k$. Clearly, we have $\sum_{k=1}^{n}p_k=1$. Without loss of
generality, we assume that pages are sorted in the decreasing order of their
popularity. That is, if $i<j$, then $p_i\geq p_j$. It is well known that
workload possesses high skewness in the sense that a small portion of data
pages receive a large fraction of requests, and the access probability
usually follows a Zipf-like distribution \cite{breslau1999web,
zipf1929relative}. Thus, we model $p_k$'s as a Zipf-like distribution.
Mathematically, we let
\begin{equation*}
  p_k=ck^{-\gamma},  \quad \gamma>0,
\end{equation*}
where $c$ is the normalized constant. We would like to emphasize that   our
model also allows other forms of distributions.

\subsection{Markov Model}

In this subsection,  we extend the
mathematical model in \cite{transient15} to  capture the dynamics of data
pages in a hybrid cache system with different architectures, and then
derive the steady-state distribution to quantify the hit ratio of each
request.

Note that we have  $n$ data pages in total in  the system, and the total
capacity of the hybrid cache is $m$.  {Without loss} of generality, we
assume that $m<n$, so only parts of data pages can be kept in the hybrid
cache.
To characterize the system state of the hybrid cache, we use a random
variable $X_{k,i}(t)$ ($k=1,2,\cdots, n$, and $i=1,2,\cdots,h$) to denote
whether page $k$ is in {list $l_i$ at }time $t$.  If yes, we let
$X_{k,i}(t)=1$ and  0 otherwise. {If page} $k$ does not exist in the hybrid
cache, i.e., $X_{k,i}(t)=0$ for $i=1,2,\cdots, h$, then page $k$ must be
stored in the secondary storage, and we let $X_{k,0}(t)=1$ in this case.

Now we capture the system state from a perspective of lists, and define
$\mathbf{Y}_{i}(t)=\{k|X_{k,i}=1\}$ ($i \in \{1,..,h\}$) as the set of pages
in {list $l_i$ at} time $t$. We have $|\mathbf{Y}_{i}(t)| \leq m_{i}$. The
process $\mathcal{
Y}^{h}(t)=(\mathbf{Y}_{1}(t),\mathbf{Y}_{2}(t),...,\mathbf{Y}_{h}(t))$
denotes the distribution of pages in the hybrid cache at time $t$. Now  the
state space of $\mathcal{Y}^{h}(t)$, which we denote as
$C_{n}(\mathbf{m})$, can be viewed as the set of all sequences of $h$ sets
$\mathbf{c}$ = $\{\mathfrak{c}_1,...,\mathfrak{c}_h\}$ with each set
$\mathfrak{c}_i$ consisting of $m_{i}$ distinct integers taken from the set
$\{1,...,n\}$.


In each time slot, only one request arrives and  triggers a state transition
accordingly. Under the independent reference model in
\S\ref{subsec:workload}, the process $\mathcal{Y}^{h}(t)$ is clearly a
Markov chain on the state space $C_{n}(\mathbf{m})$ for the  cache
replacement algorithms described in \S\ref{subsec:replacement_alg}. Now we
denote $\pi_A(\mathbf{c})$ with
$\mathbf{c}$ = $\{\mathfrak{c}_1,...,\mathfrak{c}_h\}$ as the steady-state probability of
state $\mathbf{c}$, where $A \in  \{F, L\}$ standing for the flat
architecture or the layered architecture.  We use a variable $ht_A(l_i)$
to denote the height of list $l_i$, which is defined as the number of
steps to move a data page from
{list $l_0$ to list $l_i$. Precisely}, we have

{
\begin{equation}
ht_F(l_i)\!=\!\!
\begin{cases}
i,   i = 1,...,h_N,\\
i\!-\!h_N,   i = h_N\!+\!1,...,h,\\
\end{cases} \mbox{and } ht_L(l_i) = i.
\label{eq:height}
\end{equation}
}
Now the steady-state probability $\pi_A(\mathbf{c})$ can be derived as shown
in the following theorem.

\begin{theorem}
The steady state probabilities $\pi_A(\mathbf{c})$, with $\mathbf{c}$ $\in$
$C_{n}(\mathbf{m})$, can be written as
{\footnotesize
\begin{equation}
\pi_A(\mathbf{c})=\frac{1}{Z(\mathbf{m})}\prod\nolimits_{i=1}^{h}\Big(\prod\nolimits_{j \in \mathfrak{c}_i}p_j\Big)^{ht_A(l_i)},
\label{eq:steady_state_dist}
\end{equation}
}
where
{\footnotesize
$Z(\mathbf{m})=\sum_{\mathbf{c}\in
C_{n}(\mathbf{m})}\prod_{i=1}^{h}(\prod_{j \in
\mathfrak{c}_i}p_j)^{ht_A(l_i)}$.
}
\label{theo:steady_state_dist}
\end{theorem}

\noindent{\bf Proof:}  Please refer to the \tr. \done

\noindent{\bf Remarks:} We point out that the steady-state results share the
same structure as the results in \cite{transient15} for both the flat and
layered architectures. The difference is that our model introduces a
parameter $ht_A(l_i)$, which represents the height of lists and provides the
capability of unifying the model for different architectures. In particular,
the notation $ht_A(l_i)$ (i.e., the height of lists) is an
``architecture-dependent parameter'' (i.e., its value depends on the
architecture of the hybrid system), and we include it in the analysis so as
to enhance the model's ability in analyzing different architectures.

{ According to the  probabilities $\pi_A(\mathbf{c})$,  we can calculate the
hit probability of list $l_i$ in steady state, which is denoted as
$H_{i}\!\!=\!\!\lim\limits_{t \to \infty }\sum_{k}{p_kE[X_{k,i}(t)]}$. We
also call this probability distribution {\em cache content distribution}.
Mathematically, }
\begin{equation}
H_{i}=\sum_{\mathbf{c}\in C_{n}(\mathbf{m})}\sum_{k}p_k\mathbf{1}_{\{k\in \mathfrak{c}_i\}}\pi_A(\mathbf{c}),
\label{eq:steady_state_content_dist}
\end{equation}
\vskip -5pt
%
%
where $\mathbf{1}_{\{k\in \mathfrak{c}_i\}}$ is a 0-1 variable denoting
whether page $k$ is in list $l_i$ or not.

However, it is not efficient to compute $\pi_A(\mathbf{c})$ by using the
above formula unless the cache capacity $m$ is small. In the next section,
we will introduce a mean-field approach, which can approximate the cache
content distribution very efficiently.

\subsection{Performance Metric} \label{subsec:latency}

 Recall that we focus on hybrid cache systems
consisting of both DRAM and NVM, which show very different characteristics
in access latency. To take device heterogeneity into account, we define a
latency-based performance metric to evaluate hybrid cache performance. Since
requests are processed differently under different architectures, we
distinguish the {definitions }for flat architecture and layered
architecture.

\subsubsection{Latency Model under Flat Architecture}

Suppose that at time $t$, a request arrives. To process
this request, we first access the metadata in file system to identify the
current position the request served, and there are two cases: (1) {cache} hit,
which means that the requested page is available in the hybrid cache, and (2) {cache}
miss, which means that the requested page does not exist in the hybrid cache. In the
following, we derive the access latency in the above two cases.

{At time $t$}, if cache hit happens, the service time of accessing a page
depends on which cache page is accessed. If the hit occurs in N-Cache, that
is, $\sum_{i=1}^{h_N}\sum_{k=1}^{n}p_kX_{k,i}(t)$, then the service time
includes only the  time to read a page from NVM, and we denote it as
$T_{N,r}$, where $N$ denotes N-Cache and $r$ represents read. Otherwise,
i.e., the hit occurs in D-Cache and
$\sum_{i=h_N+1}^{h}\sum_{k=1}^{n}p_kX_{k,i}(t)$, then the service time is
the time to read a page from DRAM, which we denote as $T_{D,r}$.

{ If cache miss happens, that is, $\sum_{k=1}^{n}p_kX_{k,0}(t)$, then we need
to first copy the data from the secondary storage to the destined cache
(either D-Cache or N-Cache), then serve the request from the corresponding
cache. So the service time includes the time to read a page from the
secondary storage, which we denote as $T_{S,r}$, the time to write a page to
cache, which we denote as $T_{D,w}$ for writing to D-Cache and $T_{N,w}$ for
writing to N-Cache, and the time to read a page from cache. Note that under
the flat architecture, a new data page is written to D-Cache (or N-Cache)
with probability $\alpha$ (or $1-\alpha$), so the service time in the case
of cache miss can be derived as
$\alpha(T_{S,r}+T_{D,w}+T_{D,r})+(1-\alpha)(T_{S,r}+T_{N,w}+T_{N,r})$. }

By summarizing the above two cases and {noting that
$H_i(t)=\sum_{k=1}^{n}p_kX_{k,i}(t)$}, the average service time of
processing the request at time $t$ under the flat architecture, which we
also call the average latency, can be derived as follows.

{ \vskip -8pt
\begin{eqnarray}
L_F(t)\!\!\!\!\!\!&=&\!\!\!\!\!\!E[H_0(t)]\Big(T_{S,r}\!+\!\alpha(T_{D,w}+T_{D,r})\nonumber\\
\!\!\!\!\!\!&+&\!\!\!\!\!\!(1-\alpha)(T_{N,w}+T_{N,r})\Big)+\!\!\sum_{i\neq 0}E[H_{i}(t)]T_{d(i),r},
\label{eq:flat_lantency}
\end{eqnarray}
}
where $d(i)$ is the device type of {list} $l_i$, i.e., $d(i)$ $\in$
$\{D,N,S\}$.

\subsubsection{Latency Model under Layered Architecture}
Similar to the above derivation, we can also derive the average latency
under layered architecture, while there are two differences. First, if cache
hit occurs in the highest list of N-Cache, i.e., in {list} $l_{h_N}$, then
we need to exchange this data in N-Cache with a data page in D-Cache. As a
result, we need one read from N-Cache, one write to D-Cache, as well as one
read from D-Cache and one write to N-Cache, so the total time is
$T_{N,r}+T_{D,w}+T_{D,r}+T_{N,w}$. Second, if cache miss happens, data can
only be written to N-Cache, and the service time is
$T_{S,r}+T_{N,w}+T_{N,r}$.
In summary, the average latency under the layered architecture can be
derived as: { \vskip -10pt
\begin{eqnarray}
L_L(t)\!\!&=&\!\!\!E[H_{0}(t)](T_{S,r}\!+\!T_{N,w}\!+\!T_{N,r})\nonumber\\
&&+E[H_{h_N}(t)](T_{N,r}\!+\!T_{N,w}\!+\!T_{D,r}\!+\!T_{D,w})\nonumber\\
&&+\sum\nolimits_{i\neq 0,h_N}\!\!E[H_{i}(t)]T_{d(i),r}.
\label{eq:layer_lantency}
\end{eqnarray}
\vskip -20pt
}

%
%
%


%% file: meanfield.tex
\section{Mean Field Analysis}
\label{sec:meanfield}
In this section,
we conduct mean-field analysis to approximate the cache content distribution
so as to make the computation more efficient. The rough idea of the
mean-field analysis can be stated as follows. Instead of accurately deriving
the steady-state probability distribution directly from the Markov process,
we first formulate a deterministic process defined by a set of ordinary
differential equations (ODEs),  then we show that the Markov process can be
approximated by the deterministic process, which converges to the fixed
point (i.e., mean-field limit), and finally, we use the mean field limit  to
approximate the steady-state solution of the Markov process.

\subsection{{ODEs}}
As mentioned in \cite{transient15}, the rationale of the mean-field approximation is that when
$p_{k}$ is small and the capacity of each list $m_{i}$ ($i\in
\{0,1,...,h\}$) is large, the dynamics of one particular data page becomes
independent of the hit ratio of each list, hence, its behavior can be
approximated by a time-inhomogeneous continuous-time Markov chain. As a
result, the stochastic process $\mathcal{Y}^{h}(t)$ can be approximated by a
particular deterministic process $\mathbf{x}(t)$ = \{$x_{k,i}(t)$\} ($k$ =
1, ..., $n$ and $i = 1, ..., h$).

\begin{figure}[!t]
\centering
\subfigure[Flat Architecture]
{\includegraphics[width=0.55\linewidth]{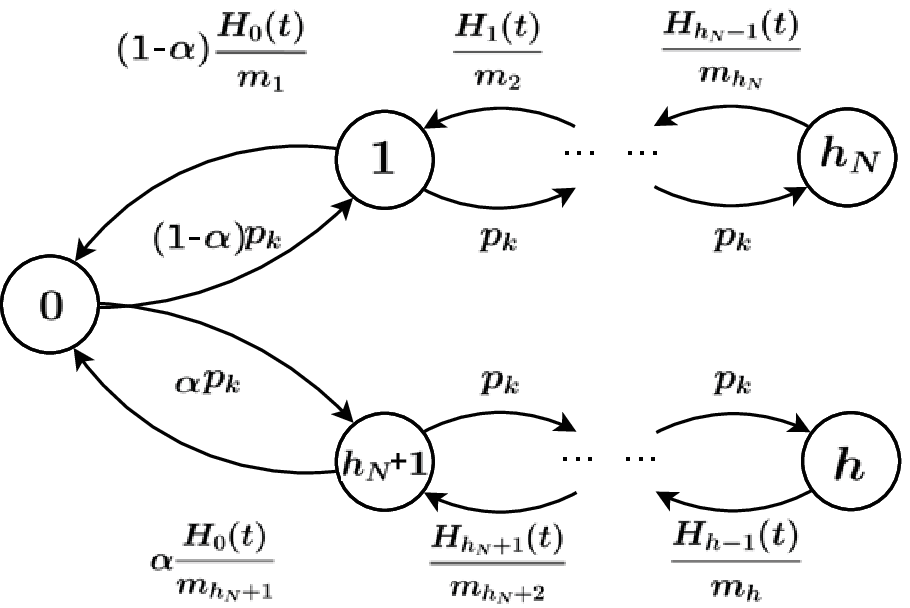}}
\subfigure[Layered Architecture]
{\includegraphics[width=0.8\linewidth]{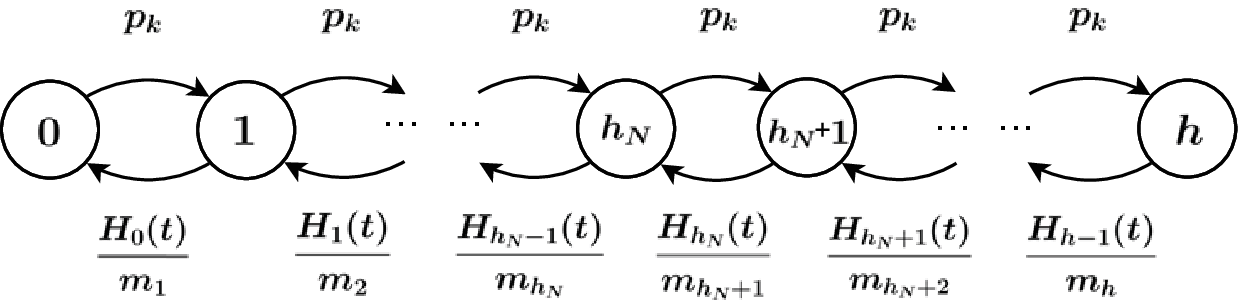}}
\vskip -5pt
\caption{State transitions of a single data page.}
\label{fig:state_transition}
\vskip -15pt
\end{figure}

To formulate the set of ODEs to define $\mathbf{x}(t)$, we first focus on
the flat architecture. According to the state transitions of a single data
page illustrated in Figure~\ref{fig:state_transition}(a), we can define
$\mathbf{x}(t)$ by using the ODEs in (\ref{eq:faltode1})-(\ref{eq:faltode5}).

\vskip 5pt {\bf \noindent Case 1:}  If $i\neq0$, $1$, $h_{N}+1$, $h, h_N$
(i.e., in middle lists): {\footnotesize
\begin{eqnarray}
\dot{x}_{k,i}(t)&=&p_{k}x_{k,i-1}(t)-\sum\nolimits_{j}p_{j}x_{j,i-1}(t)\frac{x_{k,i}(t)}{m_{i}}\nonumber\\
&+&\sum\nolimits_{j}p_{j}x_{j,i}(t)\frac{x_{k,i+1}(t)}{m_{i+1}}-p_{k}x_{k,i}(t).
\label{eq:faltode1}
\end{eqnarray}
}
{\bf\noindent Case 2:}  If $i=h$ or $i=h_N$ (i.e., in the highest list):
 {\footnotesize
\begin{eqnarray}
\dot{x}_{k,i}(t)&=&p_{k}x_{k,i-1}(t)-\sum\nolimits_{j}p_{j}x_{j,i-1}(t)\frac{x_{k,i}(t)}{m_{i}}.
\label{eq:faltode2}
\end{eqnarray}
}
{\bf\noindent Case 3:}   If  $i$ = $1$  (i.e., in the lowest list of
N-Cache):
 {\footnotesize
 \begin{eqnarray}
\dot{x}_{k,i}(t)&=&(1-\alpha) p_{k}x_{k,0}(t)-(1-\alpha)\sum\nolimits_{j}p_{j}x_{j,0}(t)\frac{x_{k,i}(t)}{m_{i}}\nonumber\\
&+&\sum\nolimits_{j}p_{j}x_{j,i}(t)\frac{x_{k,i+1}(t)}{m_{i+1}}-p_{k}x_{k,i}(t).
\label{eq:faltode3}
\end{eqnarray}
}
{\bf\noindent Case 4:}   If  $i=h_{N}+1$ (i.e., in the lowest list of
D-Cache):
 {\footnotesize
 \begin{eqnarray}
\dot{x}_{k,i}(t)\!\!\!\!&=&\!\!\!\!\alpha p_{k}x_{k,0}(t)\!-\!\alpha\sum\nolimits_{j}p_{j}x_{j,0}(t)\frac{x_{k,i}(t)}{m_{i}}\nonumber\\
&+&\sum\nolimits_{j}p_{j}x_{j,i}(t)\frac{x_{k,i+1}(t)}{m_{i+1}}-p_{k}x_{k,i}(t).
\label{eq:faltode4}
\end{eqnarray}
}
{\bf\noindent Case 5:}  If $i$ = $0$ (i.e., in the storage layer):
 {\footnotesize \begin{eqnarray}
\dot{x}_{k,0}(t)\!\!\!\!&=&\!\!\!\!(1-\alpha)\sum\nolimits_{j}p_{j}x_{j,0}(t)\frac{x_{k,1}(t)}{m_{1}}\nonumber\\
\!\!\!\!&+&\!\!\!\!\!\!\alpha\!\!\sum\nolimits_{j}p_{j}x_{j,0}(t)\frac{x_{k,h_{N}+1}(t)}{m_{h_{N}+1}}-p_{k}x_{k,0}(t).
\label{eq:faltode5}
\end{eqnarray}
\vskip -5pt
}

To illustrate the ODEs, we take~(\ref{eq:faltode1}) as an example.
First, if page $k$ is in list $i-1$ at time $t$ and it is accessed, then it
moves from list $i-1$ to $i$, and the probability is $p_{k}x_{k,i-1}(t)$.
Second, if a page in list $i-1$ is accessed, then it will exchange with a
randomly selected page in list $i$. The probability of accessing a page in
list $i-1$ is
 $\sum_{j}p_{j}x_{j,i-1}(t)$, which we denote as $H_{i-1}(t)$, and the probability of page $k$ being in list $i$ and also being selected for exchanging
 is $x_{k,i}(t)/m_{i}$. Thus, with probability $H_{i-1}(t)x_{k,i}(t)/m_{i}$, page $k$ moves from list $i$ to
list $i-1$. Third, if a page in list $i$ is accessed, then it will exchange
with a randomly selected page in list $i+1$. In this case, the probability
of  page $k$ being in list $i+1$ and moving  back to list $i$ is
$\sum_{j=1}^np_{j}x_{j,i}(t)\frac{x_{k,i+1}(t)}{m_{i+1}}$. At last, if page
$k$ is in list $i$ and  accessed, then it moves from list $i$ to list $i+1$,
and the corresponding probability is $p_{k}x_{k,i}(t)$. By summing the above
four cases, we have the ODE as in~(\ref{eq:faltode1}).

Now we consider the layered architecture, similar to the case of flat
architecture, we can also formulate the set of ODEs according to the state
transitions illustrated in Figure~\ref{fig:state_transition}(b), and the
ODEs are defined by~(\ref{eq:layerode1})-(\ref{eq:layerode2}).

\vskip 3pt

{\bf\noindent Case 1:} If $i\neq0$ (i.e., in the hybrid cache):
{\footnotesize \begin{eqnarray}
\dot{x}_{k,i}(t)\!\!\!\!&=&\!\!\!\!p_{k}x_{k,i-1}(t)-\sum\nolimits_{j}p_{j}x_{j,i-1}(t)\frac{x_{k,i}(t)}{m_{i}}\nonumber\\
\!\!\!\!&+&\!\!\!\!\!\!\mathbf{1}_{\{(i<h)\}}\!(\sum\nolimits_{j}\!p_{j}x_{j,i}(t)\frac{x_{k,i+1}(t)}{m_{i+1}}-p_{k}x_{k,i}(t)).
\label{eq:layerode1}
\end{eqnarray}
}

\noindent{\bf Case 2:}  If $i=0$ (i.e., in the storage layer):
{\footnotesize
\begin{equation}
\begin{aligned}
\dot{x}_{k,0}(t)=&\sum\nolimits_{j}p_{j}x_{j,0}(t)x_{k,1}(t)/m_{1}-p_{k}x_{k,0}(t).
\label{eq:layerode2}
\end{aligned}
\end{equation}
\vskip -3pt
}

{\noindent \bf Remarks:} We point out that the set of ODEs share
similarities with the ODEs formulated in \cite{transient15}. This is mainly
because we also divide each kind of device (DRAM or NVM)  into multiple
lists so as to explore the full design space. However, we emphasize that
with the consideration of multiple devices and different architectures, the
lists in the boundary behave in a very  different way, and so the state
transitions for boundary lists are also different.

\subsection{Fixed Point}


We derive the fixed point  of the ODEs defined by
~(\ref{eq:faltode1})-(\ref{eq:faltode5}),(\ref{eq:layerode1})-(\ref{eq:layerode2}).
The results are stated in the following theorem.

\begin{theorem}
The ODEs  have a unique fixed point, which we denote as $\pi_{k,i}$ ($k$ =
$1,...,n$ and $i$ = $0,...,h$).
{\footnotesize
\begin{equation}
\begin{aligned}
\pi_{k,i}=\frac{p_{k}^{ht_A(i)}s_{i}}{1+\sum_{j=1}^{h}p_{k}^{ht_A(j)}s_{j}},
\label{eq:fixed_point}
\end{aligned}
\end{equation}
}
where $ht_A(i)$ ($A \in \{F, L\}$) is defined in~(\ref{eq:height}),
and $(s_{1},...,s_{h})$ is the unique solution of the following equation.
{\footnotesize
\[\sum_{k=1}^{n}\frac{p_{k}^{ht_A(i)}s_{i}}{1+\sum_{j=1}^{h}p_{k}^{ht_A(j)}s_{j}}=m_{i}.\]
}
\label{theorem:fixed_point}
\end{theorem}
\vskip -10pt
\noindent {\bf Proof: } Please refer to the \tr.  \done
\vskip -10pt

\noindent {\bf Remarks:}
Note that for the layered architecture, we have $ht_A(i)=i$. By substituting
it in the above Equations in Theorem~\ref{theorem:fixed_point}, we have the
same results as in \cite{transient15}. This is because for the layered
architecture, the hybrid cache can be considered as a single unified cache
containing $h$ lists when deriving the cache content distribution. However,
we would like to emphasize that due to the device heterogeneity, the average
latency of the hybrid cache must be different from that of a single unified
cache. On the other hand, for the flat architecture, we see that in steady
state, the fixed point $\pi_{k,i}$ is independent of the parameter $\alpha$.
This implies that the hit ratio is independent of the policy of choosing
which  cache device to buffer new data. Thus, we can freely increase
$\alpha$ to cache more missed data pages in the fast-speed D-Cache so as to
achieve better overall cache performance.
In terms of the convergence, note that the stochastic process under each
architecture has the reversible property, which is the same as  Corollary 1
in \cite{transient15}, so we may also apply the method in
\cite{boudec2010stationary} to show that the process will concentrate on the
fixed point.
We also point out that the fixed-point provides
an efficient numerical method to compute the steady-state performance for
both architectures.

\subsection{{Convergence Results}}
Here, we show that we can use $\delta_i(t)=\sum_{k}x_{k,i}(t)p_k$ to
approximate $H_i(t)=\sum_{k}X_{k,i}(t)p_k$ where $x_{k,i}(t)$ is defined by
the set of ODEs.
The convergence result is stated in the following theorem.



\begin{theorem}
When $p_k\!\!\rightarrow\!\!0$ as $n\!\!\rightarrow\!\!\infty$ ($a\!=\!max_{k}p_k \rightarrow 0$) and $m_i\!\rightarrow\!\infty$, then for any $T$,
$E[\sup_{i,t\leq T}{|H_i(t)-\delta_i(t)|}] \rightarrow 0$, with initial condition $H_i(0)=\delta_i(0)$.
\label{theo:converge_to_de}
\end{theorem}

%
%

\noindent {\bf Proof: } Please refer to the \tr.  \done
\vskip -5pt

{\noindent {\bf Remarks:} Based on Theorem~\ref{theorem:fixed_point} and
Theorem~\ref{theo:converge_to_de}, we can use the fixed point
$\sum_{k}\pi_{k,i}$ (derived in~(\ref{eq:fixed_point})) to approximate the
cache content distribution $H_{i}$ (defined
in~(\ref{eq:steady_state_content_dist})), which denotes the hit probability
of list $i$ in the steady state. More importantly, it is efficient to
compute $H_{i}$ with this approximation, which makes it  feasible to further
derive the average latency of the hybrid cache. }

%
%

%% file: validation.tex
\section{Model Validation}
\label{sec:validation}

In this section,  we first validate the mean-field approximation by
comparing the hit probabilities derived from model and simulations, then we
validate our model analysis of average latency by modifying the DRAMSim2
simulator \cite{dramsim2}.

\subsection{Validation on Mean-field Approximation}
In this subsection, we validate the mean-field approximation
using the trace-based simulations by setting $m_N$ = 200, $m_D$ = 100,  $n$ = 1000,
and $p_k$ by following a Zipf-like distribution with parameter $\gamma$ = 0.8.

To validate the mean-field approximation, we use the probability of hitting
each page in each device as a metric. Note that the hit probability can be
derived from $\pi_{k,i}$. In particular, for a particular page $k$, the
probability of hitting page $k$ in N-Cache can be derived as
$\sum_{i=1}^{h_N}\pi_{k,i}$, and the  probability of hitting $k$ in D-Cache
is $\sum_{i=h_N+1}^{h}\pi_{k,i}$. For the simulation, we run 50 times and
take an average result.

\begin{figure}[!ht]
\centering
\subfigure[Flat Architecture] {\includegraphics[width=0.49\linewidth]{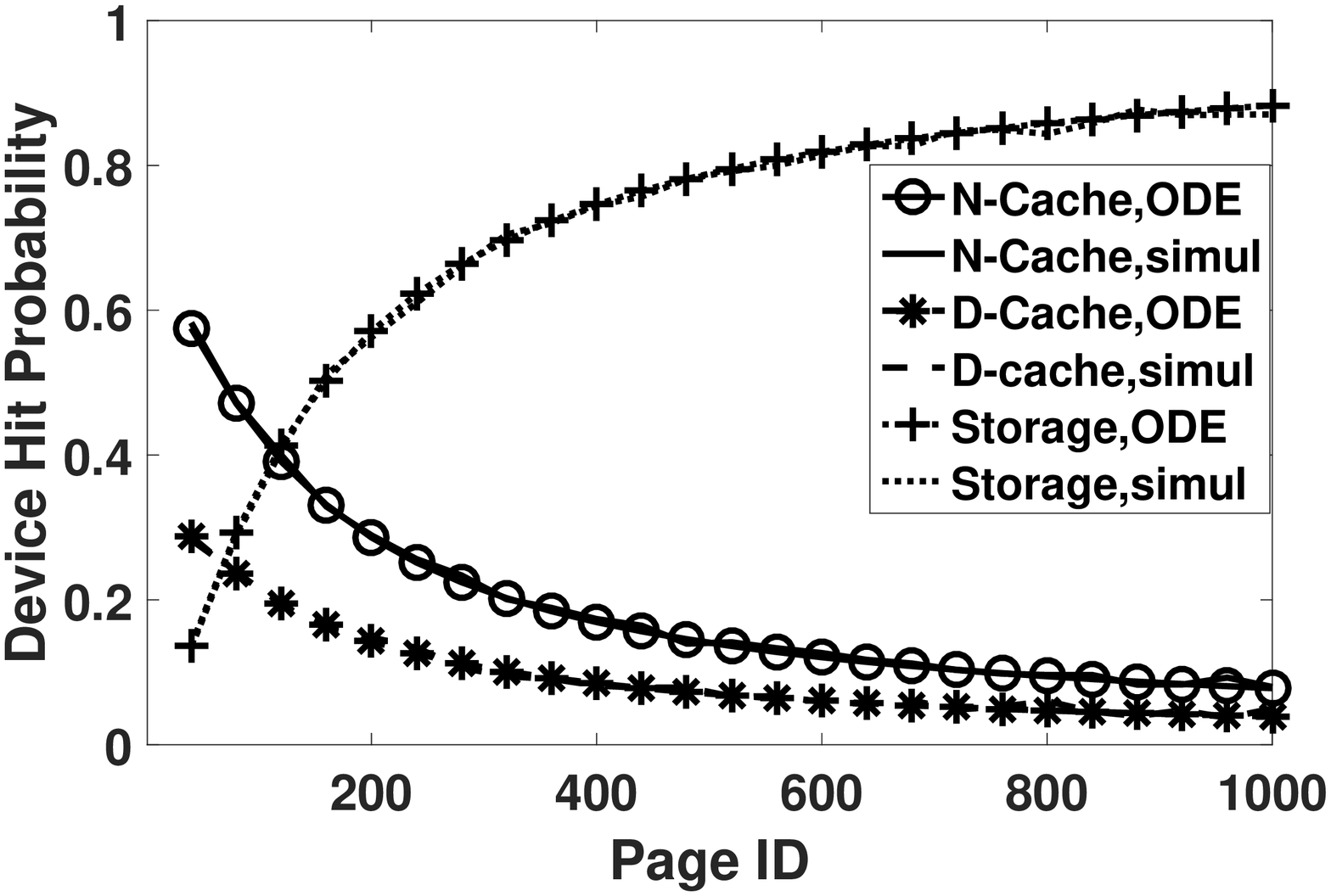}}
\subfigure[Layered Architecture] {\includegraphics[width=0.49\linewidth]{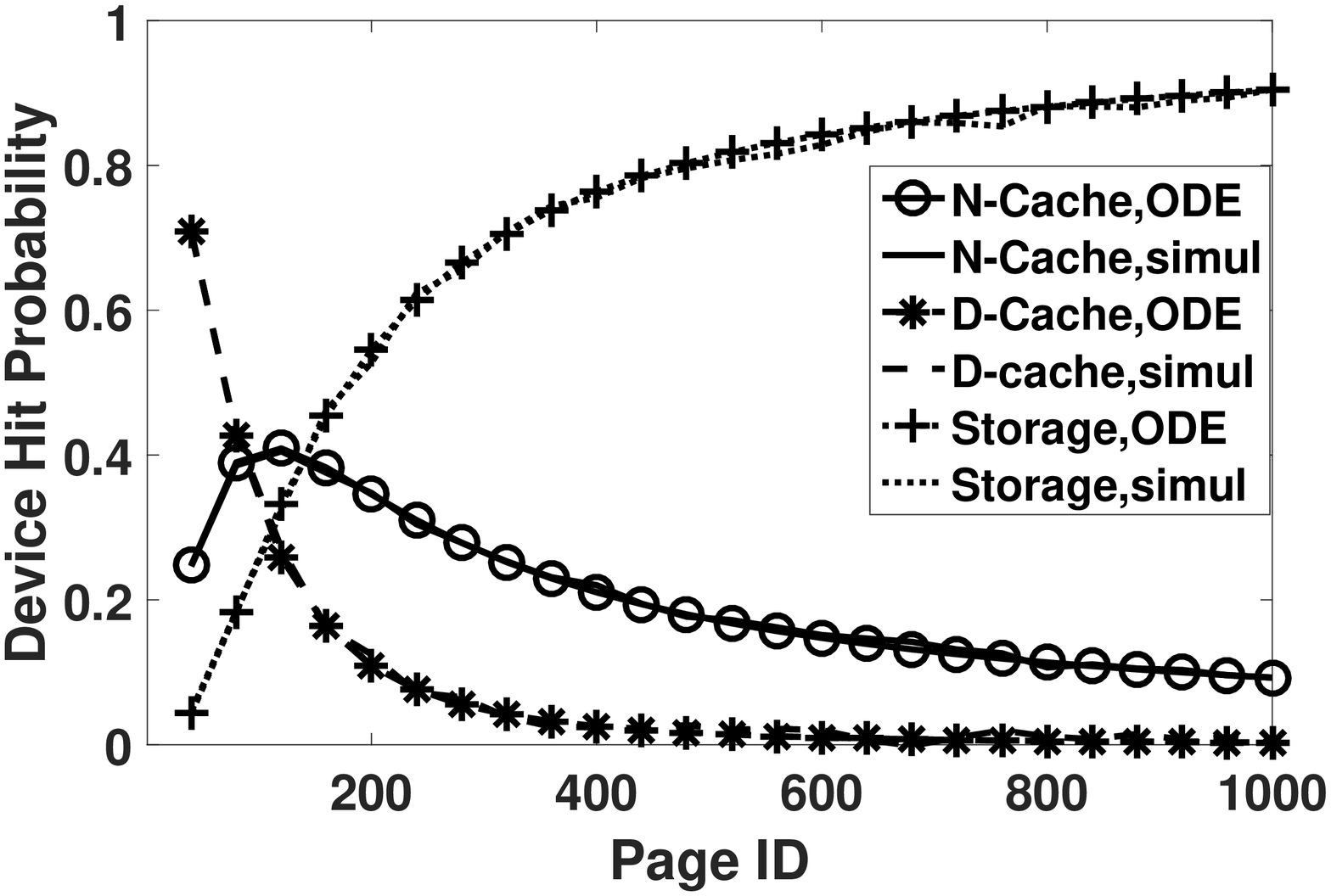}}
\vskip -5pt
\caption{Validation on mean-field approximation: The hit probability of each page
in each device.}
\label{fig:validation_mf}
\vskip -15pt
\end{figure}

Figure~\ref{fig:validation_mf} shows the model and simulation results under
the flat and layered architectures.We see that the analysis results match
well with the simulation results. In particular, even for a very small
system (e.g., $n=1000$), we can still achieve a good approximation by using
the mean-field analysis.
\begin{figure*}[!ht]
\centering
\subfigure[Flat (varying $\alpha$)] {\includegraphics[width=0.26\textwidth]{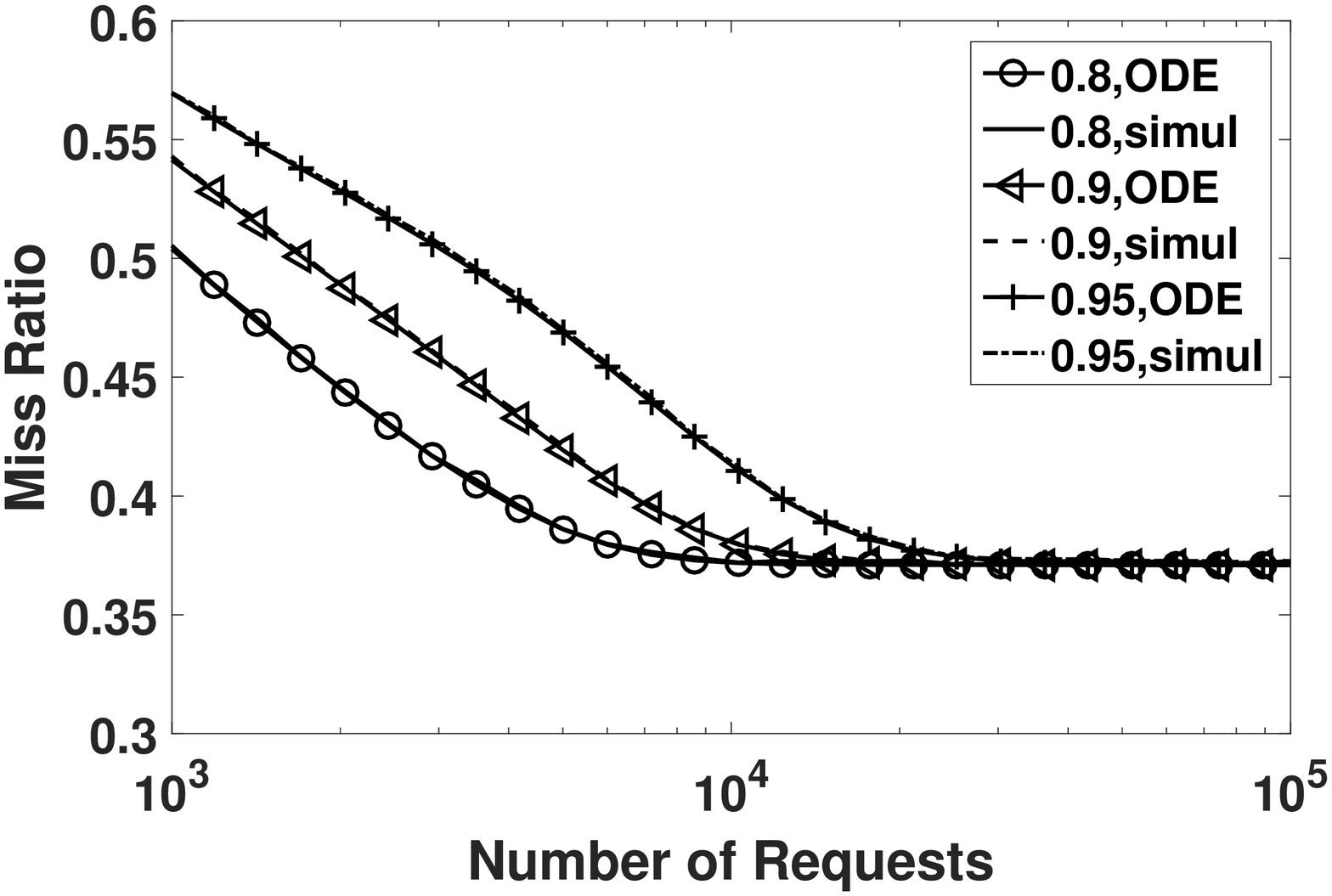}}
\hspace{5pt}
\subfigure[Flat (varying $h_N,h_D$)] {\includegraphics[width=0.26\textwidth]{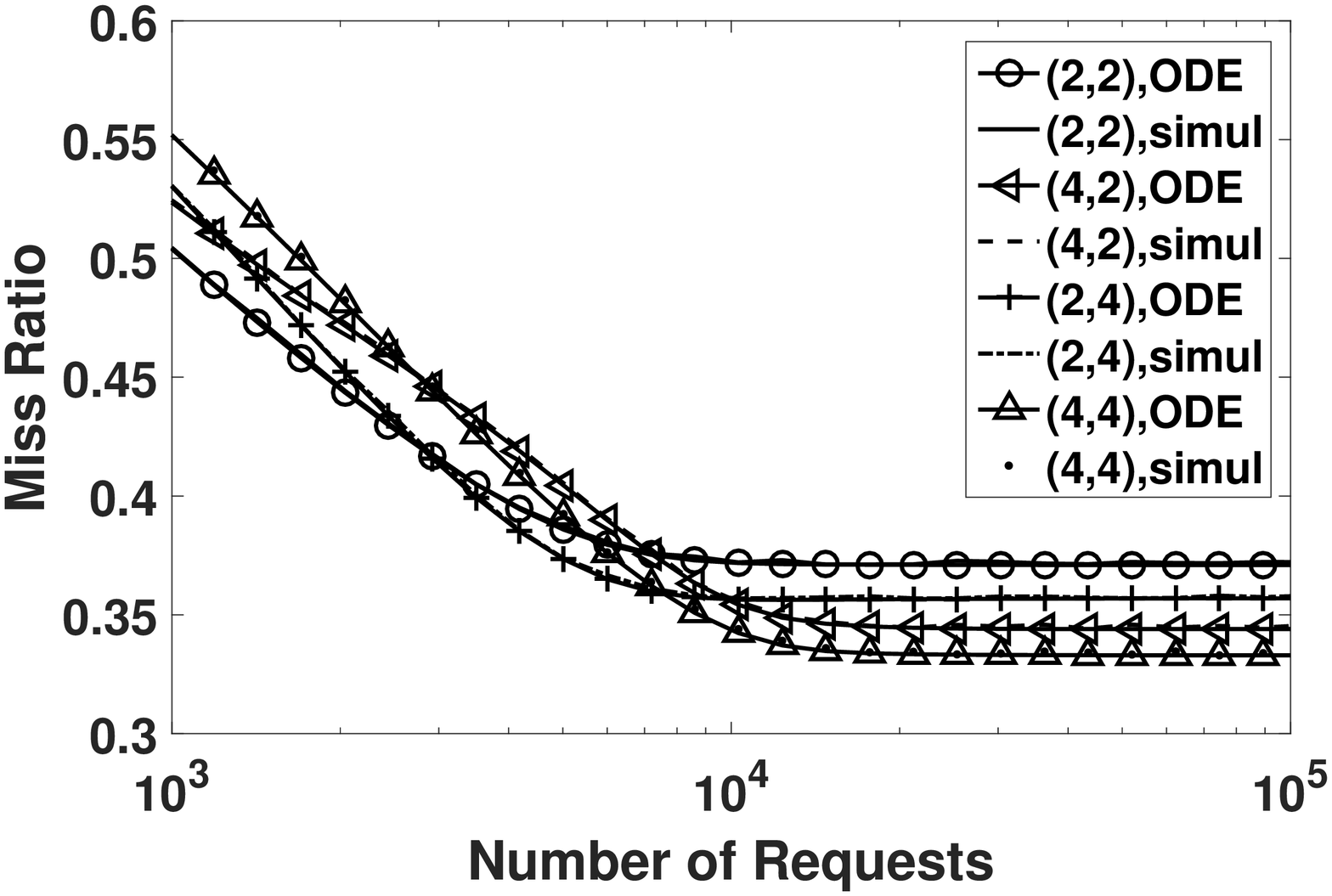}}
\hspace{5pt}
\subfigure[Layered (varying $h_N,h_D$)] {\includegraphics[width=0.26\textwidth]{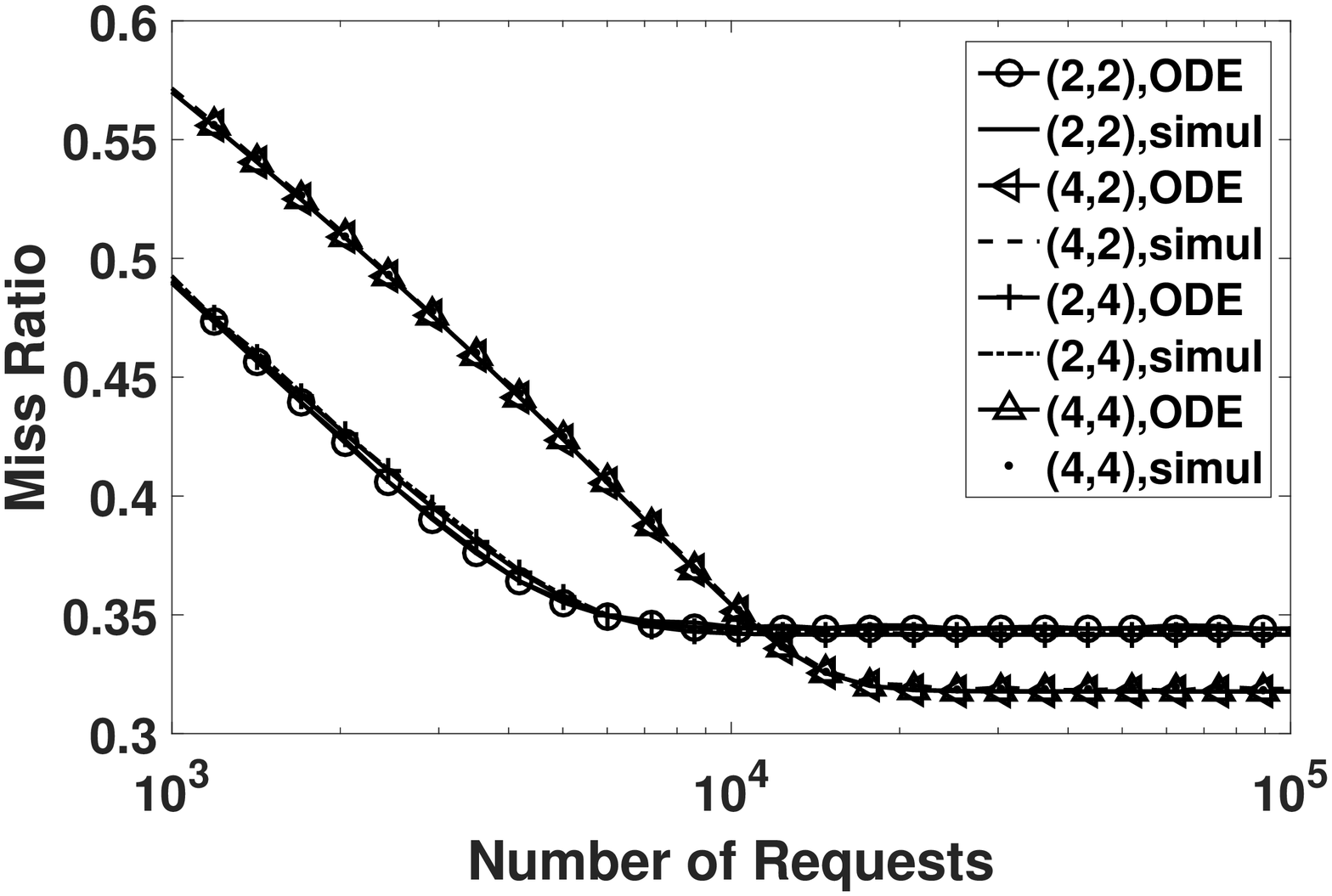}}
\vskip -8pt
\caption{Validation on mean-field approximation: Transient behaviors of the hybrid cache starting from an empty state.}
\label{fig:validation_transient}
\vskip -15pt
\end{figure*}

We further validate the mean-field approximation by considering the
transient hit probability instead of the steady-state result derived from
the mean-field limit. We use the average miss ratio of the hybrid cache over
all pages as a metric, and divide  time into small  intervals to compare the
simulation and model results in each time interval. For the model results,
since we now focus on the transient behavior, we derive the average miss
ratio directly from the ODEs in~(\ref{eq:faltode1})-(\ref{eq:faltode5})
and~(\ref{eq:layerode1})-(\ref{eq:layerode2}). Precisely, the average
miss ratio at time slot $t+1$ is computed by
$\sum_{k}p_kx_{k,0}(t+1)=\sum_{k}p_k(x_{k,0}(t)+\dot{x}_{k,0}(t))$. For
simulations, we record the position of each page after processing each
request, and then measure the average miss ratio in each time interval.

Figure~\ref{fig:validation_transient} shows the  results under different
settings by varying the parameter $\alpha$ under flat architecture
(Figure~\ref{fig:validation_transient}(a)) and varying $h_N$ and $h_D$ under
flat and layered architectures (Figure~\ref{fig:validation_transient}(b)
and~\ref{fig:validation_transient}(c)). We see that even for the
transient behavior, the mean-field model still approximates well  for small
systems. Another interesting observation is that the number of lists in each
cache device may have a big influence on the cache reactivity, which is
measured by the time to fill the cache. Precisely, if the number of lists is
set to be large, which results in a small list size, then it may need a very
long time to fill the cache. That is, the convergence  rate to the steady
state becomes small.




\subsection{Validation on Average Latency}

To validate the model analysis of average latency, we develop a hybrid cache
simulator by modifying the DRAMSim2 simulator \cite{dramsim2}, and it
includes the following modules.

{\small
\begin{itemize}
  \item \textbf{Trace Generation Module}: It generates requests with
      logical address and request starting time.
  \item \textbf{Memory Controller Module:} It manages the cache metadata
      and controls the page replacement.

  \item \textbf{D-Cache Module:} It simulates a DRAM device,serves the
      requests coming to DRAM and sends the finishing time to the
      Time Collection Module. The timing parameters of DRAM refers to \cite{lee2009architecting}.

  \item \textbf{N-Cache Module:} It simulates a NVM device and serves the
      requests coming to NVM. The default timing parameters of NVM are set
      according to  \cite{lee2009architecting},  we can vary device-level
      latency  by adjusting  the timing parameters.
  \item \textbf{Storage Module: } It simulates the access to storage
      devices by  adding a delay to the request, then sends the new time
      clock  to the Time Collection Module.
  \item \textbf{Time Collection Module:} It collects the starting time and
      finishing time of each request.
  \item \textbf{Device Performance Monitor Module:} It collects the
      average read/write latency at device level for each cache device
      (DRAM and NVM) in each time interval.
\end{itemize}
}

In our simulation, we use the  Trace Generation Module to generate requests
 according to the Zipf-like distribution, and set the
workload size $n = 3000$. We also use the Device Performance Monitor Module
to measure the  device-level latency parameters of DRAM and NVM ($T_{D,r}$,
$T_{N,r}$, $T_{D,w}$, $T_{N,w}$), and then use them as inputs to our latency
model to compute the  average latency of the hybrid cache.

We validate our model by considering different design settings, including
the system architecture, the capacity of D-Cache and N-Cache ($m_D$ and
$m_N$), and the number of lists in each cache device ($h_D$ and $h_N$). We
only show the  results  under some settings in
Table~\ref{table:validation_simulator} due to page limit. We see that the
analysis results match well with the simulation results even under the
settings of small systems, and the relative error is at most $2.87\%$. We
also run more simulations for validation by varying the timing parameters of
cache devices, results also show that our model captures the average latency
of hybrid cache accurately. We skip the results in the interest of space.

\begin{table}[!ht]
\vskip -5pt
\footnotesize
\centering
\begin{tabular}{p{0.4cm}p{0.4cm}p{0.4cm}p{0.2cm}p{0.2cm}p{1cm}p{1.2cm}p{1.2cm}}
  \hline
  \hline
Arc.& $m_N$ & $m_D$ &$h_N$ &$h_D$ &Sim.($\mu s$) &Model($\mu s$)&Rel. Err.\\
    \hline
    F&200 & 400&3&4&54.38 &55.34&1.77\%\\
    \hline
    F &200 & 400&3&3&55.67 &54.68&1.78\%\\
    \hline
    F &200 & 400&2&4&54.79 &56.21&2.59\%\\
    \hline
    F &100 & 200&3&4&69.44 &67.45&2.87\%\\
    \hline
    L&400 & 200&4&3 &49.28 &49.30 &0.04\%\\
    \hline
    L&400 & 200&3&3 &70.04 &67.54& 0.36\%\\
    \hline
    L&400 & 200&4&2 &69.22 &68.43 &1.14\%\\
    \hline
    L&300& 100&3&2 &83.56 &81.41 &2.57\%\\
    \hline
    \hline
\end{tabular}
\caption{Latency validation under different settings.}
\label{table:validation_simulator}
\vskip -25pt
\end{table}

%% file: numerical.tex
\section{Numerical results and guidelines}\label{sec:numerical}

\begin{figure*}[!ht]
\centering
\subfigure[Impact of $\alpha$]
{\includegraphics[width=0.26\textwidth]{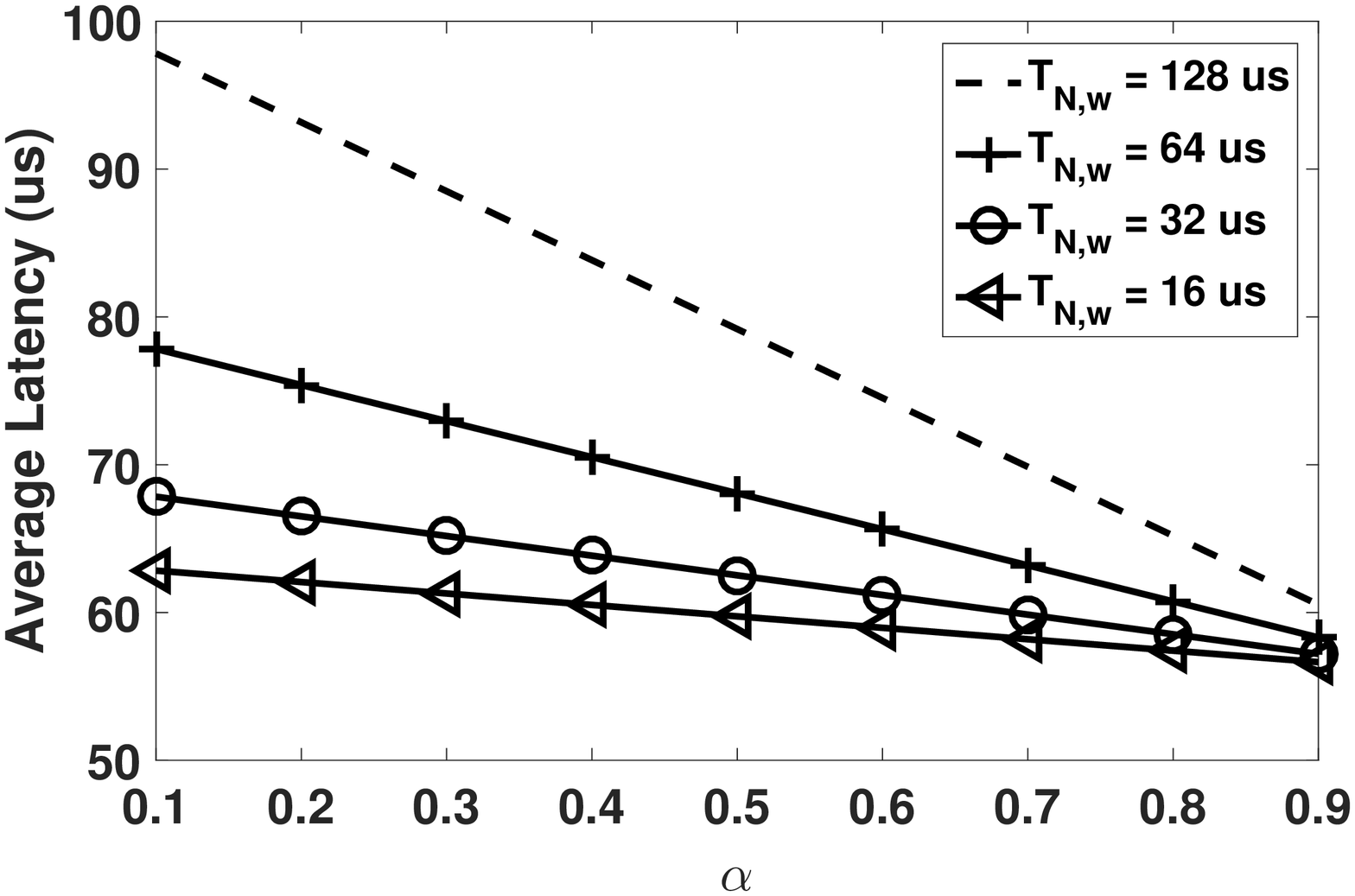}}
\hspace{5pt}
\subfigure[Impact of $h_N$]
{\includegraphics[width=0.26\textwidth]{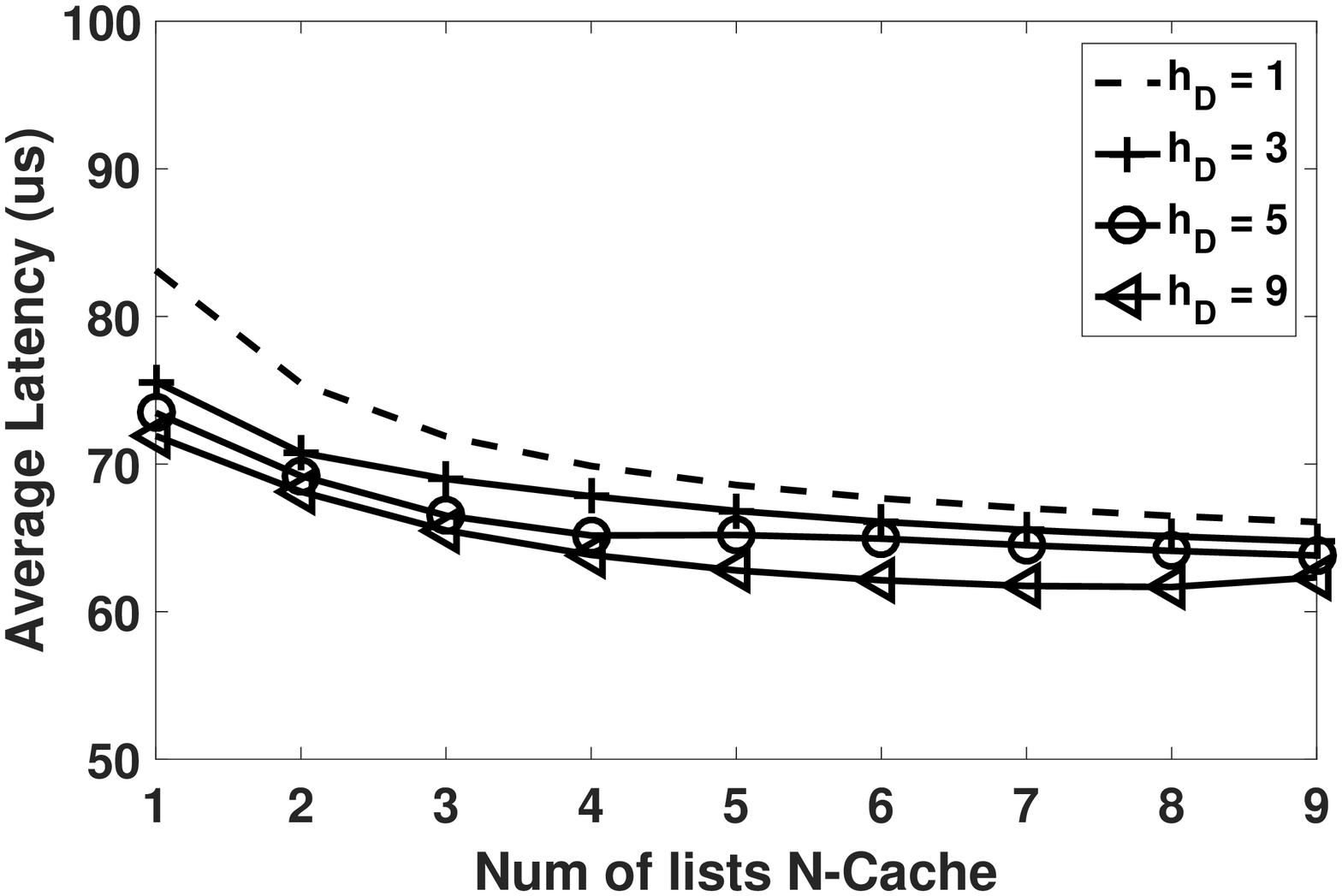}}
\hspace{5pt}
\subfigure[Impact of $h_D$]
{\includegraphics[width=0.26\textwidth]{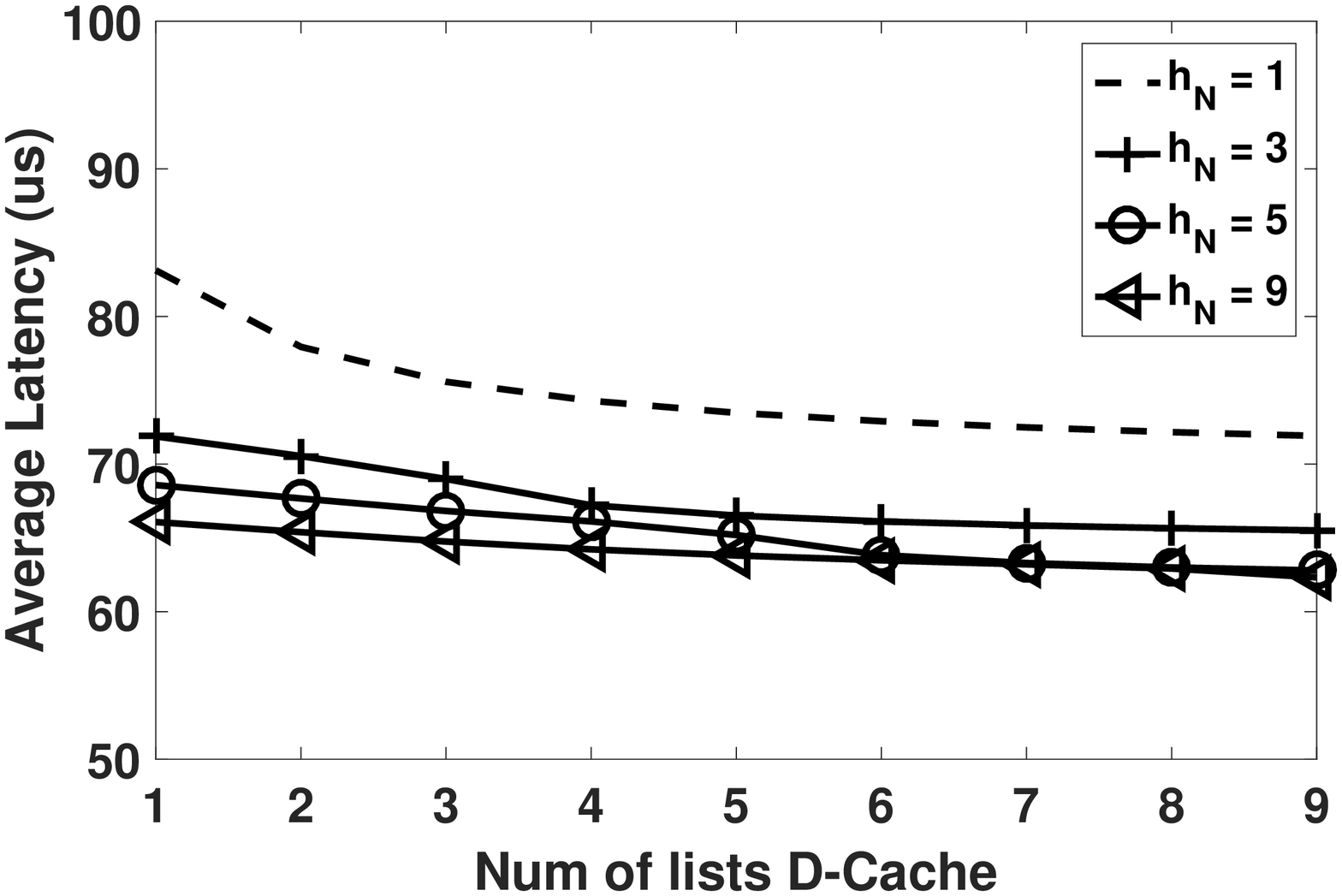}}
\vskip -5pt
\caption{Flat architecture: Impact of $\alpha$, $h_N$ and $h_D$ on average
latency of the hybrid cache ($\alpha$ denotes the probability of writing data to
D-Cache when cache miss happens, $h_N$ and $h_D$  denote the number of lists
in N-Cache and D-Cache, respectively).} \label{fig:flathphd}
\vskip -14pt
\end{figure*}

In this section, we use PCM as an example of NVM and conduct numerical analysis to study the impact of system
architecture and  design settings on hybrid cache performance so as to
understand the benefit of NVM and explore the design space of hybrid cache.
In the following, we first introduce the parameter settings  and justify
their choices, then we perform numerical analysis to study the impact of
various design choices and provide insightful guidelines.


\subsection{Parameter Settings}\label{subsec:paramter_setting}

Recall that our model takes device heterogeneity into consideration by using
a latency-based performance metric. Thus, to perform numerical analysis, we
first configure the performance parameters for  different devices.


DRAM parameters
are measured at the granularity of nanoseconds  in  practical file system
page cache environment by patching Linux kernel 4.0.2. Both  read and write
latencies are around 0.2$\mu s$, averaged over  millions of records. Note
that this latency is nearly 10$\times$ longer than that reported  in
\cite{dhiman2009pdram, ISCA09}, which is $10\sim 25ns$, this is mainly
because of the software overhead caused by file system. As PCM is not
available in the market yet, we refer to the parameter settings in
\cite{FAST14}, and let $T_{N,r}$ = 6.7$\mu s$ and $T_{N,w}$ = 128.3$\mu s$.

To set the  latency of accessing the secondary storage, we consider an
 example of networked storage application, in which the file server is
equipped with an all-flash storage system \cite{flashsystem820}. The network
parameters are based on the timing parameters  in previous work
\cite{holland2013flash}, and precisely, the network overhead for 4 KB
transmission is calculated as 41.0 $\mu s$ (8.2 $\mu s$ basic latency +
(4,096 $\times$ 8) bits $\times$ 1 $ns/$bit). Thus, the overall read time is
set as 151 $\mu s$ (110$\mu s$ file server read time + 41 $\mu s$ network
transmission overhead).

Table~\ref{tabel:para_device}  summarizes the delay parameters of different
devices used in this paper. Note that given the latency parameters in
Table~\ref{tabel:para_device} and the cache content distribution
approximated by  $\pi_{k,i}$ in~(\ref{eq:fixed_point}), the average
latency under flat  and layered architectures can be computed by
using~(\ref{eq:flat_lantency}) and~(\ref{eq:layer_lantency}),
respectively.

\begin{table}[!ht]
\footnotesize
\centering
\begin{tabular}{p{2.1cm}p{1cm}p{1cm}p{1.5cm}}
  \hline
  \hline
        & \textbf{DRAM} & \textbf{PCM} & \textbf{Storage}\\
    \hline
    \textbf{4KB R. Lat.:}   & 0.2$\mu s$ & 6.7$\mu s$ & $151\mu s$\\
    \hline
    \textbf{4KB W. Lat.:}    & 0.2$\mu s$ & 128.3$\mu s$ &  -\\
    \hline
\end{tabular}
\caption{The latency  of different devices in common setting.}
\label{tabel:para_device}
\vskip -20pt
\end{table}


\subsection{Impact of Design Choices under Flat Architecture}

In this subsection, we focus on the flat architecture, and study the impact
of various design choices by  setting $m_N$ = 15000, $m_D$ = 5000,  $n$ = 100000,
and $p_k$ by following a Zipf-like distribution with parameter $\gamma$ = 0.8.

\textbf{Impact of $\alpha$}: We first study the impact of parameter
$\alpha$, which denotes the probability of writing data to D-Cache when
cache miss happens. Note that the missed data is written to N-Cache with
probability $1-\alpha$.


Figure~\ref{fig:flathphd}(a) shows the analysis results. We see that the
average latency decreases when $\alpha$ increases. That is, if we write
missed data pages to D-Cache with higher probability, then the overall cache
performance increases, because more data will be served by the high-speed
D-Cache. However, the performance gain is limited when keep increasing PCM
performance  since the speed gap between DRAM and PCM is narrowed,
especially for the write performance. For example, if we set the PCM write
latency as $16 \mu s$, which is 8$\times$ smaller than the common setting,
then less than 10\%  improvement can be achieved when increasing $\alpha$
from 0.1 to 0.9. In the following study, we fix $\alpha$ as 0.8 under the
flat architecture.


\textbf{Impact of $h_N$ and $h_D$}: Now we study the impact of  $h_N$ and
$h_D$, which denote the number of lists in N-Cache and D-Cache,
respectively.  To decouple the dependency between $h_N$ and $h_D$, we vary
$h_N$ by fixing $h_D$ in Figure~\ref{fig:flathphd}(b), and vary $h_D$ by
fixing $h_N$ in Figure~\ref{fig:flathphd}(c). Based on the results, we have
the following observations.

\begin{itemize}
  \item Increasing the number of lists in N-Cache (i.e., $h_N$) does not
      always increase the cache performance. For example,  as shown in
      Figure~\ref{fig:flathphd}(b), when $h_D$ = 9, increasing  $h_N$
       incurs even longer latency when  $h_N$ is larger than 7. The main
       reason is that increasing the number of  lists in N-Cache may
       result in a reduction of  the overall cache miss probability of the
       hybrid cache, but it also leads to  a reduction of the cache hit
       probability of D-Cache as hot data is more likely to be trapped in
       N-Cache.

  \item The average latency decreases when using more lists in D-Cache by
      setting a larger $h_D$,  because increasing $h_D$ not only decreases
      the overall cache miss probability, but also increases the D-Cache
      hit probability. Besides, the performance gain diminishes when $h_D$
      is already large.

\end{itemize}

We also conduct analysis by varying the latency of PCM, the capacity of
N-Cache and D-Cache, and we observe the same conclusions. We do not show the
results here in the interest of space. Further considering the impact of
$h_N$ and $h_D$  on the cache reactivity (see
Figure~\ref{fig:validation_transient}), we recommend to use a large $h_D$
and a small $h_N$ under flat architecture, e.g., set $h_D$ as 4 $\sim$ 6 and
$h_N$ as 2 $\sim$ 3.


\subsection{Impact of Design Choices under Layered Architecture}

Now we focus on the layered architecture and study the impact of various
design choices. Since the major factors are $h_N$ and $h_D$ under layered
architecture, which denote the number of lists in N-Cache and D-Cache, we
also study their impact on the average latency of hybrid cache as before.

\begin{figure}[!t]
\centering
\subfigure[Impact of $h_N$] {\includegraphics[width=0.49\linewidth]{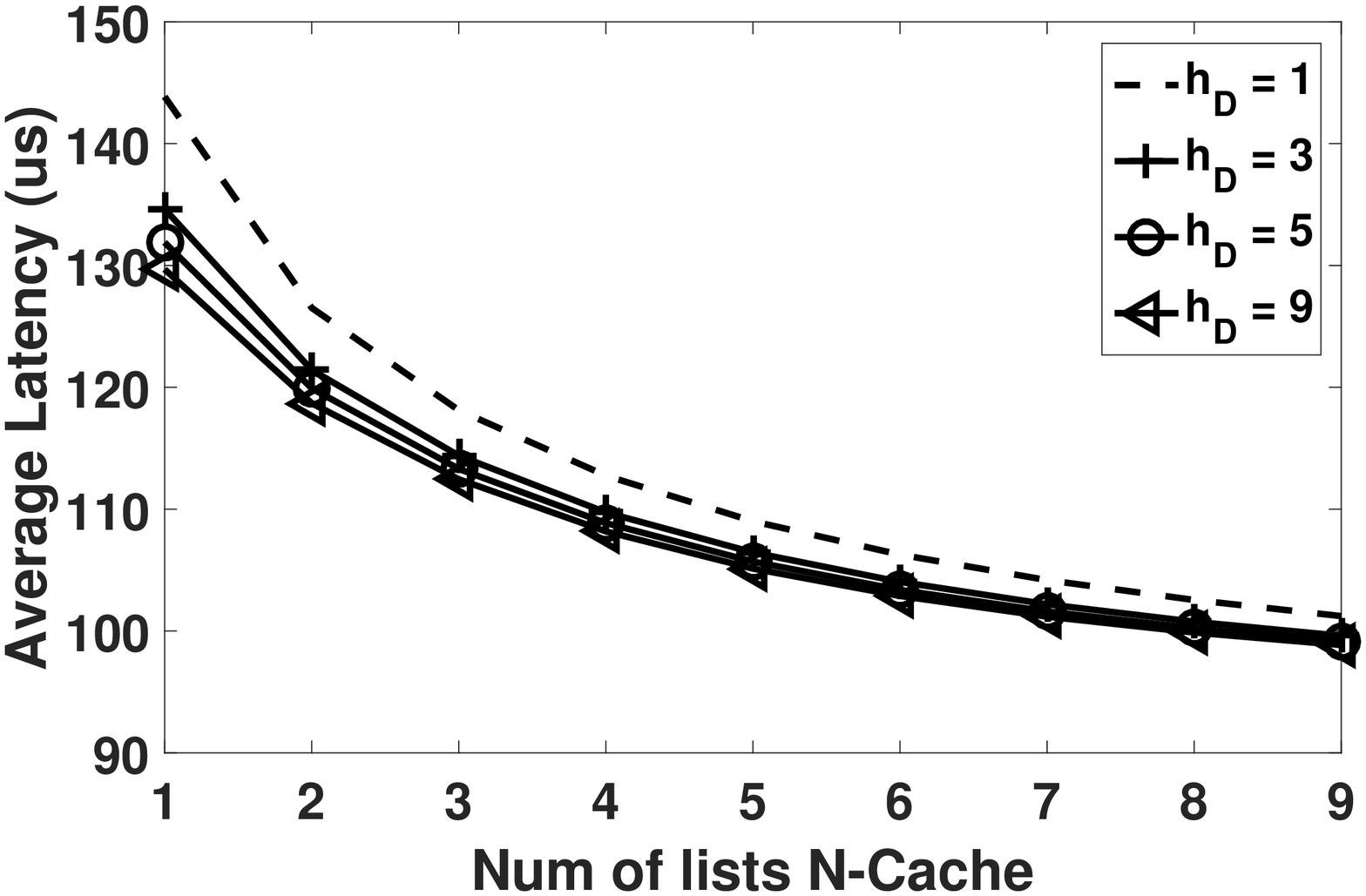}}
\subfigure[Impact of $h_D$] {\includegraphics[width=0.49\linewidth]{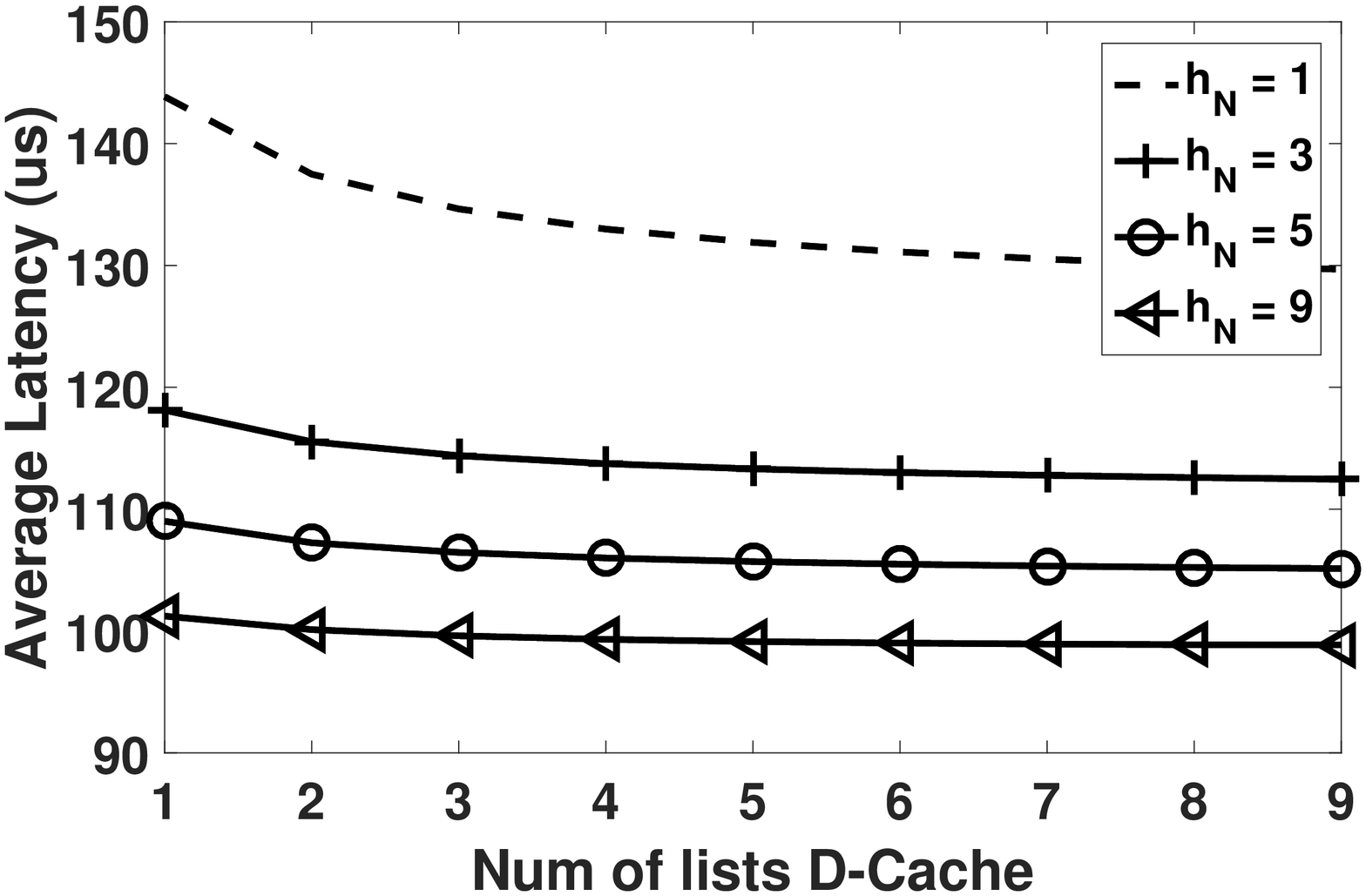}}
\vskip -5pt
\caption{Layered architecture: Impact of  $h_N$ and $h_D$  on average latency
($h_N$ and $h_D$  denote the number of lists in N-Cache and D-Cache, respectively).}
\label{fig:layeredhphd}
\vskip -15pt
\end{figure}

\textbf{Impact of $h_{N}$ and $h_{D}$: }Figure~\ref{fig:layeredhphd} shows
the analysis results, and we have the following observations.

\begin{itemize}
  \item The average latency decreases when either $h_N$ or $h_D$
      increases. That is, better cache performance can be achieved by
      adding more lists in both N-Cache and D-Cache.

  \item  The  performance improvement is more significant when adding more
      lists in N-Cache (i.e., increasing $h_N$) than increasing $h_D$. In
      particular, the  improvement is negligible when increasing $h_D$,
      especially when $h_N$ is large.
\end{itemize}

We also vary the latency of PCM and the capacity of N-Cache and D-Cache.
The results are in line with the observations. Further considering  the
impact of $h_N$ and $h_D$ on  cache reactivity (see
Figure~\ref{fig:validation_transient}), we recommend to set a large $h_N$
and a small $h_D$ under layered architecture, e.g., set $h_N$ as 4 $\sim$ 6
and $h_D$ as 2 $\sim$ 3.

\subsection{Impact of PCM Performance and Capacity}

In this subsection, we explore the performance impact and design space of
hybrid cache by varying the read and write performance of PCM, as well as
its capacity allocation. To vary the PCM capacity in hybrid cache, we fix
the total budge $C$, and adjust $m_N$ and $m_D$ by assuming that the price
of PCM is $\frac{1}{4}\times$ of that of DRAM \cite{FAST14}.

\begin{figure}[!ht]
\centering
\subfigure[Impact of $T_{N,r}$] {\includegraphics[width=0.49\linewidth]{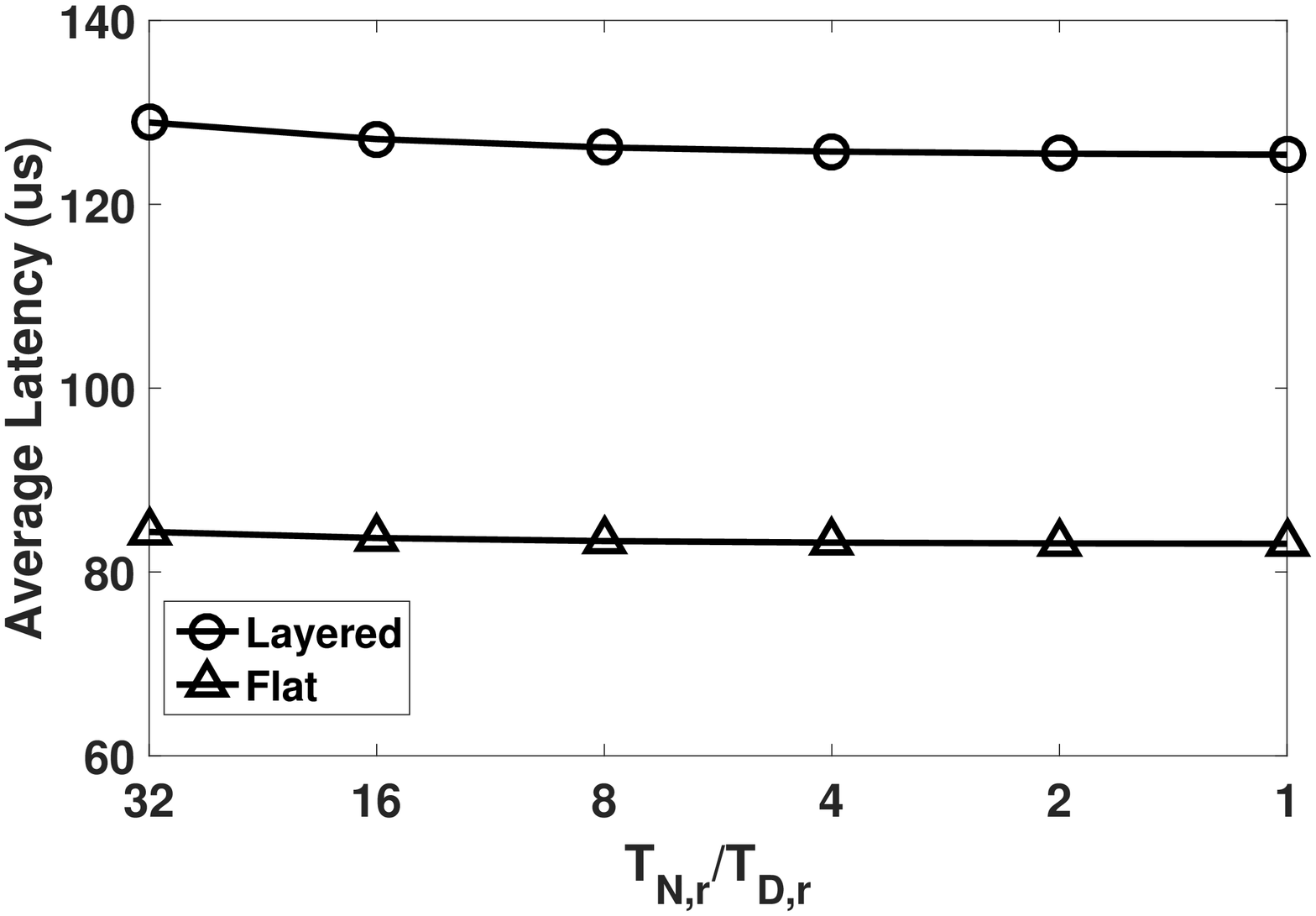}}
\subfigure[Impact of $T_{N,w}$] {\includegraphics[width=0.49\linewidth]{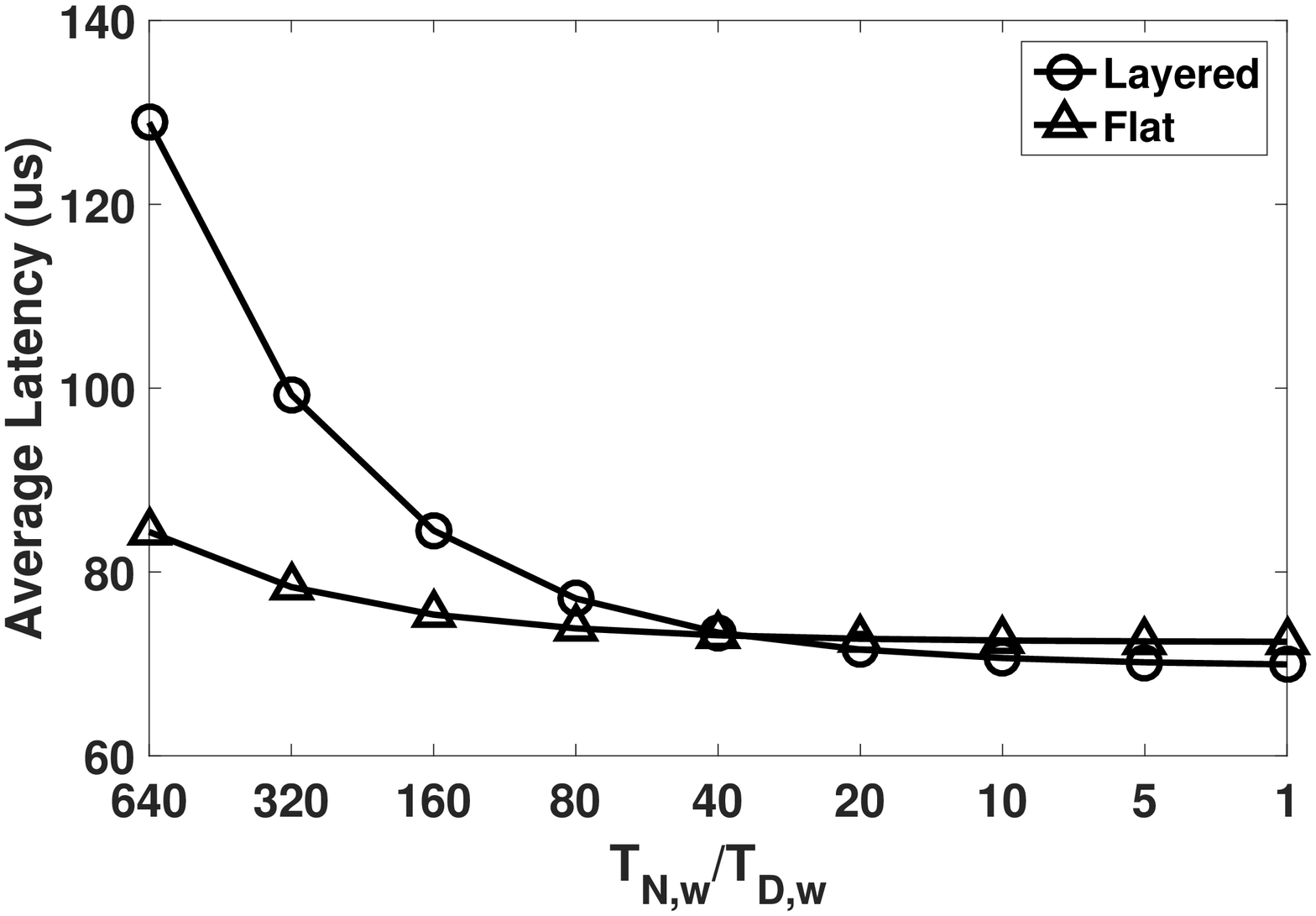}}
\vskip -5pt
\caption{Impact of PCM read/write performance ($T_{N,r}$ and $T_{N,w}$). We fix PCM capacity by setting $m_N/(m_N+m_D)=50\%$.}
\label{fig:impact_PCM_perf}
\vskip -5pt
\end{figure}

Figure~\ref{fig:impact_PCM_perf} shows the impact of PCM performance under
flat and layered architectures. In this analysis, we fix PCM capacity by
setting $m_N/(m_N+m_D)=50\%$, and we also fix the read and write performance
of DRAM ($T_{D,r}$ and $T_{D,w}$) as  the common parameters in
Table~\ref{tabel:para_device}. We change the read and write performance of
PCM by varying $T_{N,r}$ from $1\times$ to $32\times$ of $T_{D,r}$
 (see
Figure~\ref{fig:impact_PCM_perf}(a)), and varying $T_{N,w}$ from $1\times$
to $640\times$ of $T_{D,w}$ (see Figure~\ref{fig:impact_PCM_perf}(b)). Note
that in common settings, $T_{N,r}$ is $32\times$ of $T_{D,r}$ and $T_{N,w}$
is $640\times$ of $T_{D,w}$ (see Table~\ref{tabel:para_device}).

Results show that the read performance of PCM has a very small impact on the
hybrid cache performance. However, the impact of PCM write performance
$T_{N,w}$ is significant. In particular, when the write performance of PCM
is slow, the flat architecture achieves better performance than the layered
architecture, but when we increase the PCM write performance (by decreasing
$T_{N,w}$), the average latency of hybrid cache under layered architecture
drops even faster, and finally, layered architecture outperforms flat
architecture when PCM write becomes fast. Thus, choosing which architecture
in hybrid cache for better performance really depends on the PCM performance
characteristics.

\begin{figure*}[!ht]
\centering
\subfigure[Common setting  ($T_{N,w}\!\! \approx \!\!640 T_{D,w}$)] {\includegraphics[width=0.26\textwidth]{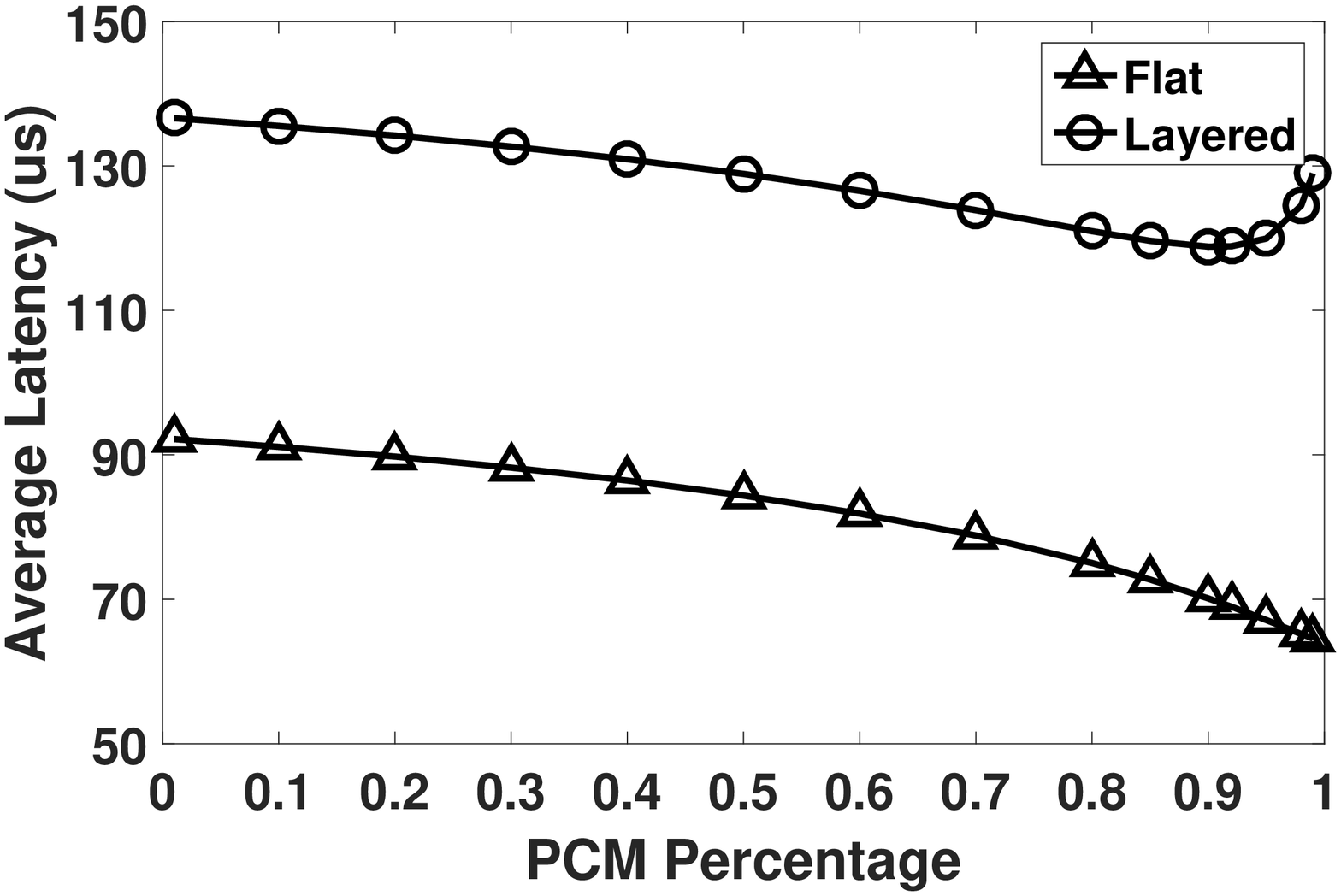}}
\subfigure[High-speed PCM ($T_{N,w} =  T_{D,w}$)] {\includegraphics[width=0.26\textwidth]{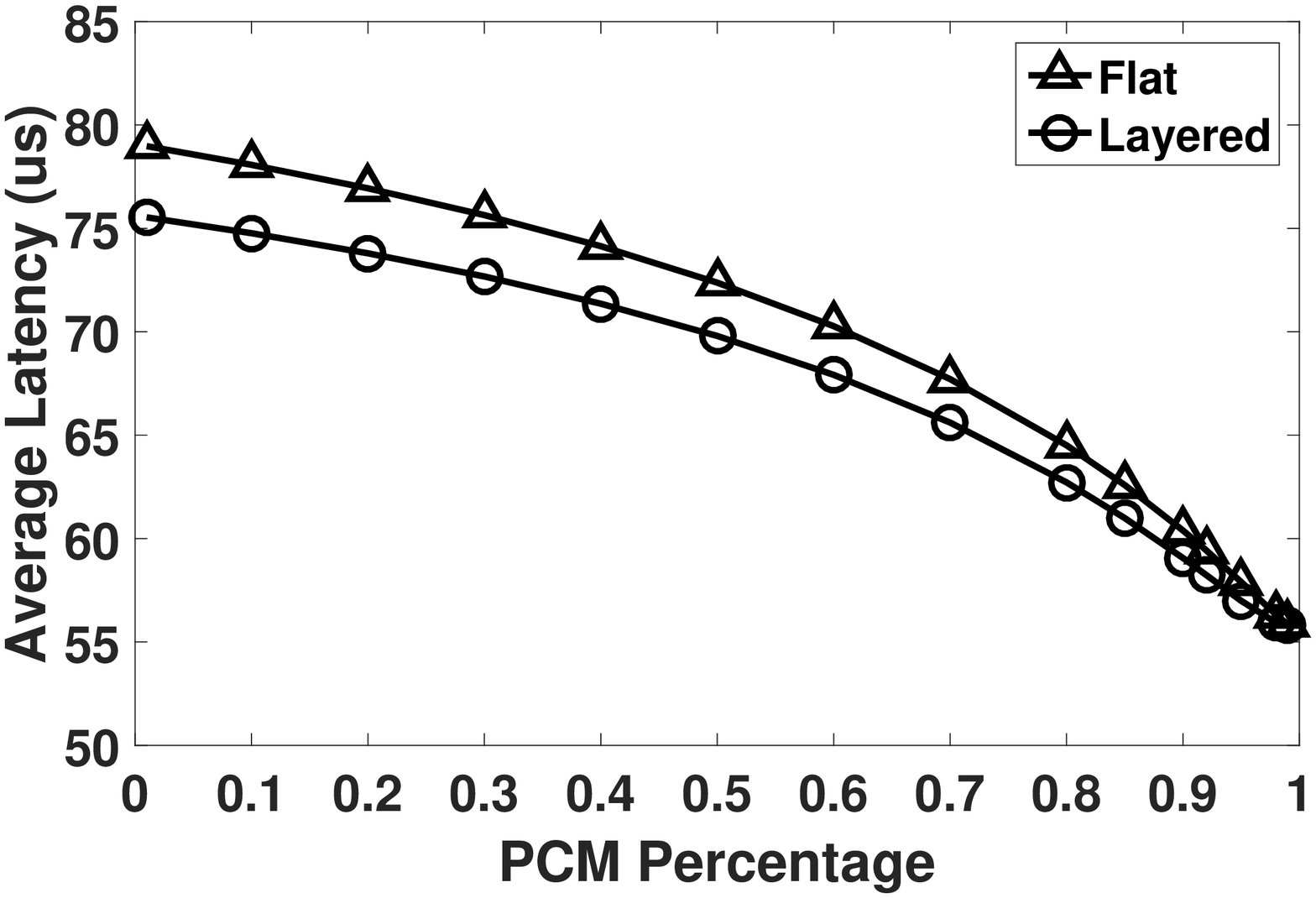}}
\subfigure[Boundary case ($T_{N,w} \approx 30 T_{D,w}$)] {\includegraphics[width=0.26\textwidth]{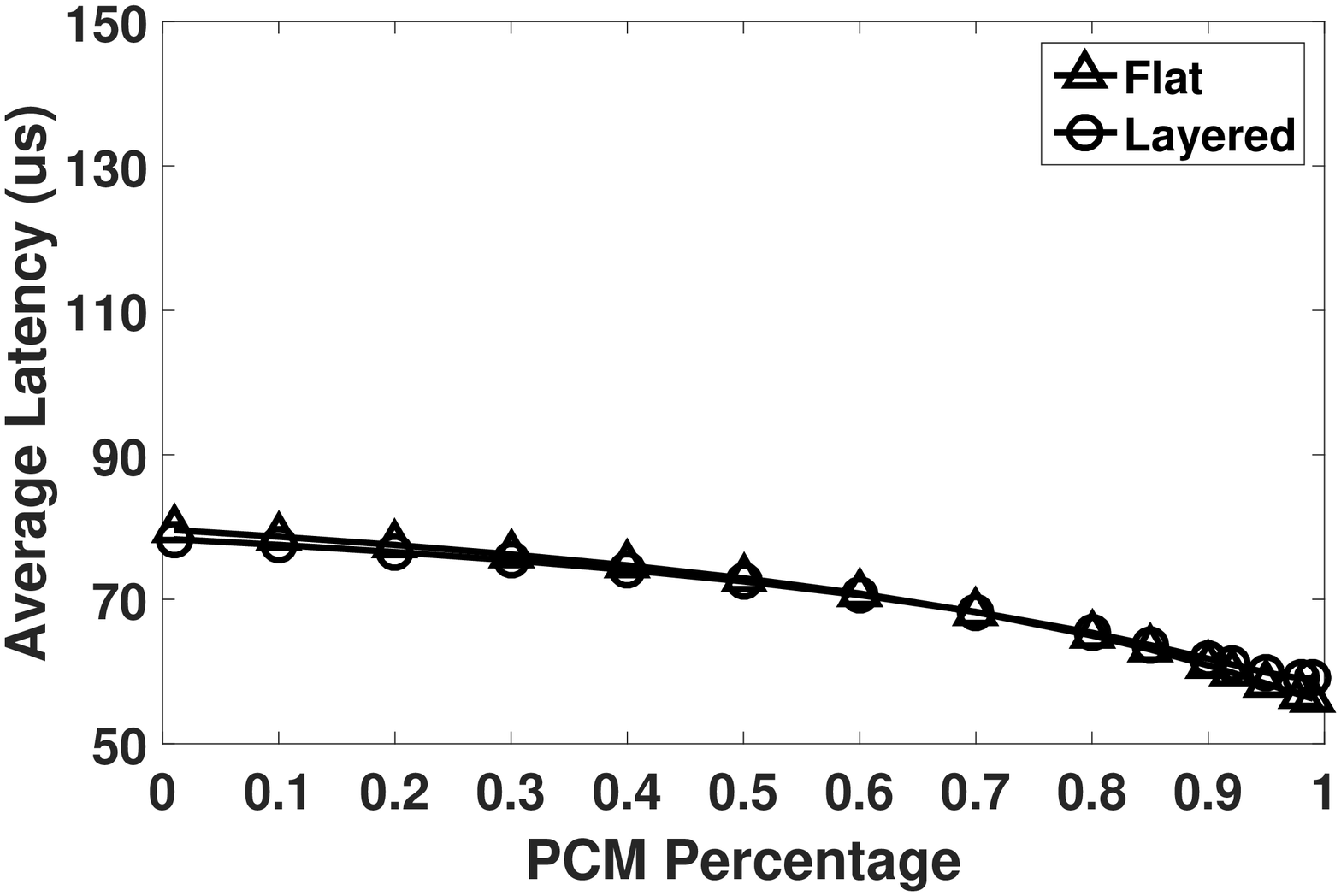}}
\vskip -5pt
\caption{{Impact of PCM capacity  in hybrid cache under different performance conditions.} }
\label{fig:impact_PCM_capacity}
\vskip -15pt
\end{figure*}

To further investigate the architectural choices of hybrid cache, we also
take into consideration the capacity allocation of different cache devices.
Results are shown in Figure~\ref{fig:impact_PCM_capacity}, in which the
horizontal axis represents the percentage of PCM cache by fixing the total
budget, and the vertical axis shows the average latency of the hybrid cache.

In Figure~\ref{fig:impact_PCM_capacity}(a), we use the common settings in
Table~\ref{tabel:para_device} to set PCM write performance (i.e., $T_{N,w}
\approx 640 T_{D,w}$). We see that if we allocate more budget for PCM, the
average latency  keeps decreasing under the flat architecture as we can have
a larger cache size. However, the result is very different for layered
architecture, in particular,  the average latency decreases first, but
begins to increase  when PCM capacity becomes very large. The main reason is
that under the layered architecture, even though we can have a large cache
by using more PCM, it may incur a lot of data migrations between DRAM and
PCM, which may incur a big overhead as PCM
 is two orders of magnitude slower than DRAM.

In terms of choosing which design between the flat and layered
architectures,
 we see that flat architecture can achieve  better
performance than layered architecture under the common setting of
performance-price ratio of PCM (as shown in
Figure~\ref{fig:impact_PCM_capacity}(a)). However, if the write performance
of PCM can have a big break through, e.g., in the extreme case where PCM
reaches the same performance as DRAM as shown in
Figure~\ref{fig:impact_PCM_capacity}(b) (i.e., $T_{N,w} = T_{D,w}$), then
layered architecture becomes  the better choice, and clearly, we do not need
to struggle with the capacity allocation problem  in this situation. To
explore the whole design space, we also seek for the boundary condition as
shown in Figure~\ref{fig:impact_PCM_capacity}(c). In general, for the common
setting of performance-price ratio  of PCM, flat architecture outperforms
layered architecture, but when the write performance of PCM improves, we may
need to switch to the layered architecture.

%% file: relatedwork.tex
\section{Related work} \label{sec:relatedwork}

In recent years, researchers are suggesting to use non-volatile devices  to
build a large memory page cache to improve system performance. For example,
researchers in \cite{Kgil2006FlashCache} and \cite{kgil2008improving}
proposed to use NAND flash memories as a page cache between DRAM  and disk
storage so as to reduce the demand of DRAM  for system memory. Lee et al. in
\cite{lee2014eliminating}  showed the potential of using a small portion of
STT-MRAM as the non-volatile buffer cache to eliminate the periodic flush
overhead caused by the volatile DRAM memory.
Especially PCM, a large body of works to study how to architect PCM in memory,
e.g., \cite{Hu2013Software,xue2011emerging,ISCA09,choi2012opamp,yoon2012row,dhiman2009pdram}.

However, recent study suggested that we should pay attention to the difference
between the material-level and the system-level performance due to the
under-developing industrial technology \cite{FAST14}. Thus, it still
necessitates a comprehensive study when incorporating NVM with DRAM  from a
system perspective.

This paper presents an analytical model to study the performance impact of
incorporating NVM in hybrid page cache  by extending the list-based model in
\cite{transient15}, and we make the following differences. First, we focus
on hybrid cache systems and consider two   system architectural designs.
Second, we take into account the device heterogeneity, and quantify the
hybrid cache performance by developing a latency model. Last, we conduct
trace-driven simulations with the DRAMSim2 simulator to validate our
analysis.

%% file: conclusion.tex
\section{Conclusions}\label{sec:conclusion}
We develop mathematical models to analyze  a hybrid cache
system so as to understand
its performance impact and design space.
We study two different
architectural designs, flat architecture and layered architecture, and
develop a latency model by taking into consideration the device
heterogeneity. We conduct trace-driven simulations with DRAMSim2 simulator
to validate our model, and perform extensive numerical analysis by
incorporating different performance characteristics and capacity ratios.
Based on our model analysis, we provide multiple guidelines on how to design
hybrid page cache so as to reach  high system throughput.

%% file: ack.tex
\section*{Acknowledgements}\label{sec:ack}
This work was supported by National Nature
Science Foundation of China (61303048 and 61379038), Anhui
Provincial Natural Science Foundation (1508085SQF214),
CCF-Tencent Open Research Fund.

%% file: app.tex
\newpage

\section*{Appendix} \label{sec:app}

\subsection{Proof of Theorem 1}

Note that, for the Layered Architecture, that is when $ht_{A}(i)=i$, [\cite{transient15}, Theorem 1] has proofed the steady state
probability.
Here, we follows the method to proof that Theorem 1 is the probability in Flat Architecture.

We use $(i,u)$ to denote the item $u$ in list $i$, that is $u\in
\mathbf{c}_{i}$, where $u$ is the item's id, and use
${\mathbf{c}}_{(i,u)\leftrightarrow (j,v)}$ to denote the new set that is
same to set $\mathbf{c}$ except that item $u$ in $\mathbf{c}_{i}$ and item
$v$ in $\mathbf{c}_{j}$ exchanged. And denote ${\mathbf{c}}_{k\rightarrow
{(i,u)}}$ as the new set that is same to set $\mathbf{c}$ except that a new
item $k$ from list $0$ is changed with item $u$ from list $i$.
For example, $\mathbf{c} =
\{\mathfrak{c}_1,\mathfrak{c}_2,\mathfrak{c}_3,\mathfrak{c}_4\} =
\{\{1,2\},\{3,4\},\{5,6\},\{7,8\}\}$, then $\mathbf{c}_{(1,2)\leftrightarrow
(2,4)}$ and $\mathbf{c}_{9\leftrightarrow (2,2)}$ can be draw as:
\[\mathbf{c}_{(1,1)\leftrightarrow (2,2)} =
\{\{1,4\},\{3,2\},\{5,6\},\{7,8\}\},\]
\[\mathbf{c}_{9\leftrightarrow (1,2)} = \{\{1,9\},\{3,4\},\{5,6\},\{7,8\}\}.\]
To prove Theorem 1, we first show lemma \ref{lemma1_theo1} as follows.
\newtheorem{lemma}{Lemma}
\begin{lemma}\label{lemma1_theo1}
For both the flat and layered architectures ($A$ = $F$ or $L$),
equation~(\ref{eq:steady_state_dist}) is equivalent to the following
equations:
$$
(*)\begin{cases}
1) \pi_A\!({\mathbf{c}}_{(i,u)\leftrightarrow (i+1,v)})/\pi_A({\mathbf{c}})\!\!=\!\!p_u/p_{v}$, $\!\!i\!\!\neq\!\!0,h_N,h,\\
2)\pi_A({\mathbf{c}}_{k\rightarrow(1,u)})/\pi_A({\mathbf{c}}) = p_{k}/p_{u}\\
3)\pi_A({\mathbf{c}}_{k\rightarrow(h_N+1,u)})/\pi_A({\mathbf{c}}) = p_{k}/p_{u}
\end{cases}$$
\end {lemma}


\noindent{\bf Proof of Lemma~\ref{lemma1_theo1}: } First, we can see that
Equation~(\ref{eq:steady_state_dist}) $\Rightarrow$(*) holds clearly.
   In order to prove  (*) $\Rightarrow$~(\ref{eq:steady_state_dist}),
  we label all the states $\mathbf{c}$ in $C_{n}(\mathbf{m})$ as $\mathbf{c}_1,\mathbf{c}_2,...,\mathbf{c}_{|C_{n}(\mathbf{m})|}$.
  For each page $k$ in state $\mathbf{c}_j$, we define $I_{k}(\mathbf{c}_j)$ as the height of the list that contains page $k$ in
  state $\mathbf{c}_j$, e.g. in state $\mathbf{c}_j$, assuming list $l_z$ contains page $k$,
  then $I_{k}(\mathbf{c}_j)$ can be drawn as $ht_A(l_z)$.
  We normalize each probability with respect to $\mathbf{c}_1$.
  Using the (*), we draw the ratio of other state $\mathbf{c}_j$'s steady state probabilities $\pi_A(\mathbf{c}_j)$ $(j\in C_{n}(\mathbf{m}), j\neq i)$ to that of $\mathbf{c}_1$:
\begin{equation}
\pi_A(\mathbf{c}_j)/\pi_A(\mathbf{c}_1)=\Big(\prod\nolimits_{k = 1}^{n} {p_k}^{I_{k}(\mathbf{c}_j)-I_{k}(\mathbf{c}_1)}\Big)
\label{eq:steadycjc1_theo1}
\end{equation}
  By using that $\sum_{j=1}^{|C_{n}(\mathbf{m})|}\pi_A(\mathbf{c}_j) = 1$, as all the steady state probabilities sums as 1, this yeilds

\begin{equation*}
\Big(1+\sum_{j=2}^{|C_{n}(\mathbf{m})|}\prod\nolimits_{k = 1}^{n} {p_k}^{I_{k}(\mathbf{c}_j)-I_{k}(\mathbf{c}_1)} \Big)\pi_A(\mathbf{c}_1) = 1.
\label{eq:steadyc_theo1}
\end{equation*}

So we get
\begin{equation*}
\pi_A(\mathbf{c}_1)=1/\Big(1+\sum_{j=2}^{|C_{n}(\mathbf{m})|}\prod\nolimits_{k = 1}^{n}{p_k}^{I_{k}(\mathbf{c}_j)-I_{k}(\mathbf{c}_1)} \Big).
\end{equation*}

By multiple both the numerator and denominator in the left hand side by $\prod\nolimits_{k = 1}^{n}{p_k}^{I_{k}(\mathbf{c}_1)}$,
we get
\begin{equation*}
\pi_A(\mathbf{c}_1)=\prod\nolimits_{k = 1}^{n}{p_k}^{(I_{k}(\mathbf{c}_1)}/\sum_{j=1}^{|C_{n}(\mathbf{m})|}\prod\nolimits_{k = 1}^{n}{p_k}^{I_{k}(\mathbf{c}_j)},
\label{eq:steadyc1_theo1}
\end{equation*}

this implies that
\begin{equation}
\pi_A(\mathbf{c}_1)=\frac{1}{Z(\mathbf{m})}\prod\nolimits_{i=1}^{h}\Big(\prod\nolimits_{j \in {\mathfrak{c}_1}_i}p_j\Big)^{ht_A(l_i)},
\label{eq:steady_state_c}
\end{equation}
By using (\ref{eq:steadycjc1_theo1}), we can draw other states's steady
state probabilities $\pi_A(\mathbf{c}_j)$. So far, we proof that (*)
$\Rightarrow$~(\ref{eq:steady_state_dist}) holds. \done

To prove Theorem 1,  we start with the flat architecture. In flat
architecture, we first derive the transition probabilities between state
$\mathbf{c}$ and another state, which are as follows.
\begin{itemize}
  \item The probability that state $\mathbf{c}$ is transited to another state can be expressed as:
  \begin{eqnarray}
\pi_F(\mathbf{c})\!\!\!\!\!\!\!\!\!&&(1-\sum_{j\in \mathfrak{c}_{h_N}}p_j-\sum_{j\in \mathfrak{c}_{h}}p_j),
\label{eq:transition_prob1}
\end{eqnarray}
noting that this transition will happen unless there is a hit on the highest list in D-Cache or N-Cache.
  \item Then we express the probability that the other states are transited back to state $\mathbf{c}$.
Using the notations above, the probability that the other states are transited back to state $\mathbf{c}$
can be expressed as
\begin{eqnarray}
\!\!\!\!\!\!\!\!&&\!\!\!\!\!\!\!\!(1-\alpha)\sum_{k \notin \mathfrak{c}_1,...,\mathfrak{c}_h}\sum_{u\in \mathfrak{c}_1}\pi_F({\mathbf{c}}_{k\rightarrow (1,u)})p_{u}/m_{1}\\
\!\!\!\!\!\!\!\!&&\!\!\!\!\!\!\!\!+\alpha\!\!\!\sum_{k \notin \mathfrak{c}_1,...,\mathfrak{c}_h}\sum_{u\in \mathfrak{c}_{{h_N+1}}}\!\!\!\pi_F({\mathbf{c}}_{k\rightarrow (h_N+1,u)})p_{u}/m_{h_N+1}\nonumber\\
\!\!\!\!\!\!\!\!&&\!\!\!\!\!\!\!\!+\!\!\!\sum_{i \neq 0,h_N,h}\sum_{u\in\mathfrak{c}_{i}}\!\sum_{v \in \mathfrak{c}_{{i+1}}}\!\!\!\pi_F({\mathbf{c}}_{(i,u)\leftrightarrow (i+1,v)})p_{v}/m_{i+1}.\nonumber
\label{eq:transition_prob2}
\end{eqnarray}
\end{itemize}

Reaching the steady state means that the probability that state $\mathbf{c}$
is transited to another state (i.e., (\ref{eq:transition_prob1})) equals the probability that other states are
transited back to state $\mathbf{c}$ (i.e., (\ref{eq:transition_prob2})). We can express
the global balance equation of state $\mathbf{c}$ as follows,
\begin{eqnarray}
&&\pi_F(\mathbf{c})\!(1-\sum_{j\in \mathfrak{c}_{h_N}}p_j-\sum_{j\in \mathfrak{c}_{h}}p_j)\\
\!\!\!\!\!&=&\!\!\!\!\!(1-\alpha)\!\!\!\sum_{k \notin \mathfrak{c}_1,...,\mathfrak{c}_h}\!\!\sum_{u\in \mathfrak{c}_{1}}\pi_F({\mathbf{c}}_{k\rightarrow (1,u)})p_{u}/m_{1}\nonumber\\
\!\!\!\!\!&+&\!\!\!\!\!\alpha\sum_{k \notin \mathfrak{c}_1,...,\mathfrak{c}_h}\!\!\sum_{u\in \mathfrak{c}_{{h_N+1}}}\!\!\!\pi_F({\mathbf{c}}_{k\rightarrow (h_N+1,u)})p_{u}/m_{h_N+1}\nonumber\\
\!\!\!\!\!&+&\!\!\!\!\!\sum_{i \neq 0,h_N,h}\sum_{u\in \mathfrak{c}_{i}}\sum_{v\in \mathfrak{c}_{{i+1}}}\!\!\!\pi_F({\mathbf{c}}_{(i,u)\leftrightarrow (i+1,v)})p_{v}/m_{i+1}.\nonumber
\label{eq:steady_prob}
\end{eqnarray}

By plugging (*) into (\ref{eq:steady_prob}), we draw that

\begin{eqnarray*}
1-\sum_{j \in \mathbf{c}_{{h_N}}}p_{j}-\sum_{j\in \mathbf{c}_{h}}p_{j} = \sum_{k \notin \mathbf{c}}p_k+\sum_{i=1}^{h}{\sum_{u\in \mathbf{c}_{i}}p_{u}},
\label{eq:proof_them1}
\end{eqnarray*}
which clearly holds as $\sum p_k = 1$, denoting that (*) is the steady state
probabilities. By using lemma \ref{lemma1_theo1}, we proof that Theorem 1
holds for the flat architecture. We see that in steady state, the steady
state of each state in flat architecture (i.e., $\pi_F(\mathbf{c})$) is
independent of the parameter $\alpha$. \done

\subsection{Proof of Theorem 2}
For the Layered Architecture, that is when $ht_{A}(i)=i$, \cite{transient15} Theorem 7 has proofed that the ODEs have a fixed point and the fixed point is unique.
Here, we follows the method to show the uniqueness in Flat Architecture.

Note that the equation  has the same structure with a birth-and-death
process. For list $i \in \{1,...,h_{N}\}$, we have
\[x_{k,i}=\frac{p_{k}^im_{1}m_{2}...m_{i}}{H_{0}H_{1}...H_{i-1}}x_{k,0},i \in \{0,1,...,h_{N}\}.\]

Noting that the list $i \in \{h_{N}+1,...,h\}$ is the $(i-h_{N})$-th list in
D-Cache, so we have
\[x_{k,i}=\frac{p_{k}^{i-h_{N}}m_{h_{N}+1}...m_{i}}{H_{0}H_{h_{N}+1}...H_{i-1}}x_{k,0},i \in \{h_{N}+1,...,h\}\]
To simplify the above equation, we define $s$ by letting
$$s_{i}=
\begin{cases}
\frac{m_{1}m_{2}...m_{i}}{H_{0}H_{1}...H_{i-1}}, & \text{if $i$ $\in$ $\{1,...,h_{N}\}$},\\
\frac{m_{h_{N}+1}...m_{i}}{H_{0}H_{h_{N}+1}...H_{i-1}}, & \text{if $i$ $\in$ $\{h_{N}+1,...,h\}$}.
\end{cases}$$

Clearly, we have $\sum_{j} x_{k,j} = 1$, which implies that
$x_{k,i}=\frac{p_{k}^{ht_{F}(i)}s_{i}}{1+\sum_{j=1}^{h_{N}}p_{k}^js_{j}+\sum_{j=h_{N}+1}^{h}p_{k}^{j-h_{N}}s_{j}}$.
By using that $\sum_{k=1}^{n} x_{k,j} = m_{j}$, we can have
\[m_{i}=\sum_{k=1}^{n}
x_{k,i}=\sum_{k=1}^{n}\frac{p_{k}^{ht_{F}(i)}s_{i}}{1+\sum_{j=1}^{h_{N}}p_{k}^js_{j}+\sum_{j=h_{N}+1}^{h}p_{k}^{j-h_{N}}s_{j}}.\]

For a vector $\vec{s}=(s_1,s_2,...,s_h)$,$i\in\{1,...,h\}$, where each $s_i >0 $.
Define $D_{i}(\vec{s})$,$i\in\{1,...,h\}$ by
\[D_{i}(\vec{s})=\sum_{k=1}^{n}\frac{p_{k}^{ht_{F}(i)}s_{i}}{1+\sum_{j=1}^{h_{N}}p_{k}^js_{j}+\sum_{j=h_{N}+1}^{h}p_{k}^{j-h_{N}}s_{j}}.\]

Define a vector $\vec{s}_{i}(y)$ to denote the vector that all the elements equal to $\mathbf{s}$ except that the $i-th$
element is $y$. $D_{i}(\vec{s}_{i}(y))=m_i$ has a unique solution that we denotes as $G_{i}(\vec{s})$. Hence $G(\vec{s})$
is a vector that calculated by $\vec{s}$.
Since
\[D_{i}(\vec{s})=\sum_{k=1}^{n}\frac{p_{k}^{ht_{F}(i)}}{1/s_i+\sum_{j=1}^{h_{N}}p_{k}^js_{j}/s_{i}+\sum_{j=h_{N}+1}^{h}p_{k}^{j-h_{N}}s_{j}/s_{i}}.\]
it implies that $D_{i}(\vec{s})$ is decreasing in $s_j$ when $j\neq i$, which implies that $G_i(\vec{s})$ is increasing in $\vec{s}$.
We define the sequence $\vec{s}(t)$ by $\vec{s}(0) = (0,0,...,0)$ and $\vec{s}(t+1) = G(\vec{s}(t))$.

Moreover, for any $t$, $D_{i}(\vec{s}(t)) \leq m_i$, which implies that
\[n-\sum_{i=1}^{h}m_i\leq n-\sum_{i=i}^{h}D_{i}(\vec{s}(t))\]
\[=n-\sum_{k=1}^{n}\frac{\sum_{i=1}^{h_{N}}p_{k}^is^t_{i}+\sum_{i=h_{N}+1}^{h}p_{k}^{i-h_{N}}s^t_{i}}{1+\sum_{j=1}^{h_{N}}p_{k}^js^t_{j}+\sum_{j=h_{N}+1}^{h}p_{k}^{j-h_{N}}s^t_{j}}\]
\[=\sum_{k=1}^{n}\frac{1}{1+\sum_{j=1}^{h_{N}}p_{k}^js^t_{j}+\sum_{j=h_{N}+1}^{h}p_{k}^{j-h_{N}}s^t_{j}}\].

Therefore, we can conclude that the sequence $\vec{s}(t)$ is bounded, that
is the sequence is increasing and it finally converges to a fixed point $G$.

Now we prove the uniqueness. Let $\lambda > 1$ and multiply all the elements
in vector $\mathbf{s}$ by $\lambda$, we have:
\[D_i({\lambda\vec{s}}) = \sum_{k=1}^{n}\frac{p_{k}^{ht_{F}(i)}\lambda s_{i}}{1+\sum_{j=1}^{h_{N}}p_{k}^j\lambda s_{j}+\sum_{j=h_{N}+1}^{h}p_{k}^{j-h_{N}}\lambda s_{j}}\]
\[= \sum_{k=1}^{n}\frac{p_{k}^{ht_{F}(i)} s_{i}}{1/\lambda+\sum_{j=1}^{h_{N}}p_{k}^j s_{j}+\sum_{j=h_{N}+1}^{h}p_{k}^{j-h_{N}} s_{j}} >  D_{i}(\vec{s})\]
This implies that $G_{i}(\lambda\vec{s})< \lambda G_{i}(\vec{s})$, which by
[\cite{yates1995framework},Theorem 2] proof the uniqueness. \done

\subsection{Proof of Theorem 3}
For the Layered Architecture, that is when $ht_{A}(i)=i$, [\cite{transient15} Theorem 7] has proofed the validity of the approximation.
Here, we follows the method to show validity of using $\vec{\delta}(t)$ to approximate $\vec{H}(t)$ in Flat Architecture.

Let ${\vec{H}}(t)$ be the vector that describe the hit probability in each list at time $t$, where each element $H_{i}(t)=\sum_{k}p_kX_{k,i}(t)$ is the hit probability of list $i$. Recall that $X_{k,i}(t)$ denotes whether
item $k$ is in list $i$ at time $t$. If yes, $X_{k,i}(t) = 1$ and 0 otherwise. Hence $0\leqslant H_{i}(t)\leq1$.

Let ${\vec{\delta}}(t)$ be the vector that consists of $\delta_{i}(t)$, each element $\delta_{i}(t)$ is defined by $\delta_{i}(t)=\sum_{k}p_kx_{k,i}(t)$, where $x_{k,i}(t)$ is the unique solution of ODEs in (\ref{eq:faltode1})-(\ref{eq:faltode5}), with initial conditions $x_{k,i}(0)=X_{k,i}(0)$.
So the sequence $\vec{\delta}(t)$, $t = 0,1,...$ describes the deterministic process.
By varying the initial conditions, we get series of sequence. At any time of the sequence, it is a vector, denoted
by $\vec{\delta}$. Let $\mathbb{H}$ be the name space of
vector $\vec{\delta}$. And we define two norms on any $\vec{\delta}$ in $\mathbb{H}$:${\|\vec{\delta}\|}_{\infty}
= \max_{i}|\delta_{i}|$, ${\|\vec{\delta}\|}^2_{\infty} =
\max_{i}|\delta_{i}|^2$.

Under the Flat Architecture, we define the function $f$ on $\delta$ as
follows. \vskip 5px {\bf \noindent Case 1:}  If $i\neq0$, $1$, $h_{N}+1$,
$h, h_N$ (i.e., in middle lists): {\small
\begin{eqnarray}
f_{i}(\vec{\delta})&=&p_k\delta_{i-1}-\frac{\delta_{i-1}\delta_{i}}{m_i}\nonumber\\
&+&\frac{\delta_{i}\delta_{i+1}}{m_{i+1}}-p_k\delta_{i}.
\label{eq:app_faltode1}
\end{eqnarray}
}

{\bf\noindent Case 2:}  If $i=h$ or $i=h_N$ (i.e., in the highest list):
 {\small
\begin{eqnarray}
f_{i}(\vec{\delta})&=&p_k\delta_{i-1}-\frac{\delta_{i}\delta_{i-1}}{m_i}.
\label{eq:app_faltode2}
\end{eqnarray}
}

{\bf\noindent Case 3:}   If  $i$ = $1$  (i.e., in the lowest list of
N-Cache):
 {\small
 \begin{eqnarray}
f_i(\vec{\delta})&=&(1-\alpha)p_k\delta_{i-1}-(1-\alpha)\frac{\delta_{i}\delta_{i-1}}{m_i}\nonumber\\
&+&\frac{\delta_{i}\delta_{i+1}}{m_{i+1}}-p_k\delta_{i}.
\label{eq:app_faltode3}
\end{eqnarray}
}

{\bf\noindent Case 4:}   If  $i=h_{N}+1$ (i.e., in the lowest list of
D-Cache):
 {\small
 \begin{eqnarray}
f_i(\vec{\delta})&=&\alpha p_k\delta_{i-1}-\alpha \frac{\delta_{i-1}\delta_{i}}{m_i}\nonumber\\
&+&\frac{\delta_{i}\delta_{i+1}}{m_{i+1}}-p_k\delta_{i}.
\label{eq:app_faltode4}
\end{eqnarray}
}

{\bf\noindent Case 5:}  If $i$ = $0$ (i.e., in the storage layer):
{\small\begin{eqnarray}
f_i(\vec{\delta})\!\!\!\!&=&\!\!\!\!(1-\alpha)\frac{\delta_0\delta_1}{m_{1}}\nonumber\\
\!&+&\!\alpha\!\frac{{\delta}_0\delta_{h_{N}+1}}{m_{h_{N}+1}}-p_{k}\delta_0.
\label{eq:app_faltode5}
\end{eqnarray}
}

Now we first prove that the following four lemmas hold.
\begin{lemma}
\label{app_lemma1}
$f(\vec{H}(t))$ is the average variation of $\vec{H}(t)$, i.e.,
\[E[\vec{H}(t+1)-\vec{H}(t)|{\mathcal{F}}_{t}]=f(\vec{H}(t)).\]
\end {lemma}

\begin{lemma}
\label{app_lemma2}
The second moment of the variation of $\vec{H}(t)$ is bounded is bounded:
\[E[{\|\vec{H}(t\!+\!1)-\vec{H}(t)\|}^2_{\infty}|{\mathcal{F}}_{t}]\leq 2a^2.\]
\end{lemma}

\begin{lemma}
\label{app_lemma3}
There exists a constant $L$ which is independent with $p_{k}$'s and
$m_{i}$'s, such that the function $f$ is Lipschitz-continuous of constant
$L(a+b)$ on ${\mathbf{X}}$, that is: for all  ${\delta}^{'}$ and ${\delta}^{''}$ in $\mathbb{H}$,
\[{\|f(\vec{{\delta}^{'}})-f(\vec{{\delta}^{''}})\|}_{\infty}\leq L(a+b){\|\vec{{\delta}^{'}}-\vec{{\delta}^{''}}\|}_{\infty}\]
where $a =max_{k}p_{k}$ and
$b=max_{i}(1/m_{i})$.
\end {lemma}

\begin{lemma}
\label{app_lemma4}
With initial conditions ${\delta}_{i}(0)=E[H_{i}(0)]$, then we can draw that:
\[\vec{\delta}(t)=\vec{H}(0)+\sum_{s=0}^{t-1}f(\vec{\delta}(s)).\]
\end {lemma}

\vskip  5px \noindent{\bf Proof of Lemma \ref{app_lemma1}: } Take  case 1 as
an example. If $i\neq0$, $1$, $h_{N}+1$, $h, h_N$ (i.e., in middle lists),
at time $t$, two types of events can modify the value of $H_{i}$:

\begin{itemize}
  \item If at time $t$, an item in list $i-1$ is requested, denoted as $k$, $k\in\{1,...,n\}$, and it exchanges with an item
from list $i$, denoted as $j$, $j\in\{1,...,n\}$ . The average variation of $H_{i}$ due to these events is:
{\small
\begin{eqnarray}
&&\sum_{k,j}\frac{X_{k,i-1}(t)X_{j,i}(t)p_k}{m_i}(p_k-p_j)\nonumber\\
&=&\sum_{k,j}\frac{X_{k,i-1}(t)X_{j,i}(t)p_k}{m_i}p_k\nonumber\\
&&-\sum_{k,j}\frac{X_{k,i-1}(t)X_{j,i}(t)p_k}{m_i}p_j\nonumber\\
&=&p_k\sum_{j}\frac{X_{j,i}(t)}{m_i}\sum_{k}X_{k,i-1}(t)p_k\nonumber\\
&&-\sum_{k}X_{k,i-1}(t)p_k\sum_{j}\frac{X_{j,i}(t)p_j}{m_i}\nonumber\\
&=&p_kH_{i-1}(t)-\frac{H_{i-1}(t)H_{i}(t)}{m_i}
\label{eq:app_proof3_1}
\end{eqnarray}
}
  \item If at time $t$, item $k$ is requested in list $i$, and it exchanges with an item
$j$ from list $i+1$. The average variation of $H_{i}$ due to these events is:

{\small
\begin{eqnarray}
&&\sum_{k,j}\frac{X_{k,i}(t)X_{j,i+1}(t)p_k}{m_{i+1}}(p_j-p_k)\nonumber\\
&=&\sum_{k,j}\frac{X_{k,i}(t)X_{j,i+1}(t)p_k}{m_{i+1}}p_j\nonumber\\
&&-\sum_{k,j}\frac{X_{k,i}(t)X_{j,i+1}(t)p_k}{m_{i+1}}p_k\nonumber\\
&=&\sum_{j}X_{j,i+1}(t)p_j\sum_{k}X_{k,i}(t)p_k/m_{i+1}\nonumber\\
&&-p_k\sum_{j}X_{j,i+1}(t)/m_{i+1}\sum_{k}X_{k,i}(t)p_k\nonumber\\
&=&\frac{H_{i+1}(t)H_{i}(t)}{m_{i+1}}-p_kH_{i}(t)
\label{eq:app_proof3_2}
\end{eqnarray}
}

\end{itemize}
By summing the two terms, we have for $i\neq0,1,h_{N},h$:
\[E[H_i(t+1)-H_i(t)|{\mathcal{F}}_{t}]=f_i(\vec{H}(t))\]
Note that $i$ is chosen randomly, we have
\[E[\vec{H}(t+1)-\vec{H}(t)|{\mathcal{F}}_{t}]=f(\vec{H}(t))\]
By summing up the cases of all the values of $i$, we can prove Lemma 2.
\done

\vskip  5px \noindent{\bf Proof of Lemma~\ref{app_lemma2}: }The second
moment of the variation of $\vec{H}(t)$ can be derived as follows.
{\small\begin{eqnarray}
&&E[(H_i(t+1)-H_i(t))^2|{\mathcal{F}}_{t}]\!\!\!\!\nonumber\\
&=&\sum_{k,j}X_{k,i-1}(t)X_{j,i}(t)(p_k-p_j)^2p_k/m_i\nonumber\\
\!&+&\!\sum_{k,j}X_{k,i}(t)X_{j,i+1}(t)(p_j-p_k)^2p_k/m_{i+1}
\label{eq:proof_theorem3_lemma3_1}
\end{eqnarray}
}

Since $0<p_i,p_k\leq \max_{k}p_k = a$, we have
$E[(H_i(t+1)-H_i(t))^2|{\mathcal{F}}_{t}]$ is less than:

{\small\begin{eqnarray}
&& \frac{\sum_{k,j}X_{k,i-1}(t)X_{j,i}(t)p_ka^2}{m_i}+\frac{\sum_{k,j}X_{k,i}(t)X_{j,i+1}(t)p_ka^2}{m_{i+1}}\nonumber\\
&=&(H_{i-1}(t)+H_{i}(t))a^2
\label{eq:proof_theorem3_lemma3_1_1}
\end{eqnarray}
}

This shows that:
{\small\begin{eqnarray}
&&E[{\|\vec{H}(t+1)-\vec{H}(t)||}^2_{\infty}|{\mathcal{F}}_{t}]\!\!\!\!\nonumber\\
&=&E[\sup_{i}(H_i(t+1)-H_i(t))^2|{\mathcal{F}}_{t}]\nonumber\\
&\leq&E[\sum_{i}(H_i(t+1)-H_i(t))^2|{\mathcal{F}}_{t}]\nonumber\\
&\leq&\sum_{i}(H_{i-1}(t)+H_{i}(t))a^2\nonumber\\
&\leq&2a^2
\label{eq:proof_theorem3_lemma3_2}
\end{eqnarray}
}

Thus, the second moment of the variation of $\vec{H}(t)$ is bounded. \done

 \vskip 5px \noindent{\bf Proof of Lemma~\ref{app_lemma3}: } First we take case 1 as an example, that is $i\neq0$, $1$, $h_{N}+1$, $h, h_N$
(i.e., in middle lists): {\small
\begin{eqnarray}
f_{i}(\delta)&=&p_k\delta_{i-1}-\frac{\delta_{i-1}\delta_{i}}{m_i}\nonumber\\
&+&\frac{\delta_{i}\delta_{i+1}}{m_{i+1}}-p_k\delta_{i}\nonumber
\label{eq:app_faltode1}
\end{eqnarray}
}

We split  $f_{i}(\vec{\delta})$ into four parts and denote each part as
$g_{i}(\vec{\delta})$ for ease of presentation. We then show that each part
is Lipschitz-continuous individually. We denote $\vec{\delta^{'}}$ and
$\vec{\delta^{''}}$ as two vectors which are chosen randomly from
$\mathbb{H}$.

\begin{itemize}
  \item Part 1: $g_{i}(\vec{\delta})=p_k\delta_{i-1}$. For any $i$,
      $i\neq0$, $1$, $h_{N}+1$, $h, h_N$:
  \[|g_{i}(\delta^{'}_{i})-g_{i}(\delta^{''}_{i})|=|p_k(\delta^{'}_{i}-\delta^{''}_{i})|\leq p_k|\delta^{'}_{i}-\delta^{''}_{i}|\]
  \[\leq a{\|\vec{{\delta}^{'}}-\vec{{\delta}^{''}}\|}_{\infty}\]
  Hence ${\|g(\vec{{\delta}^{'}})-g(\vec{{\delta}^{''}})\|}_{\infty} \leq
  {a\|\vec{{\delta}^{'}}-\vec{{\delta}^{''}}\|}_{\infty}$.
  \item Part 2: $g_{i}(\vec{\delta})=-\frac{\delta_{i-1}\delta_{i}}{m_i}$.
      For any $i$, $i\neq0$, $1$, $h_{N}+1$, $h, h_N$:
  \[|g_{i}(\delta^{'}_{i})-g_{i}(\delta^{''}_{i})|=|\frac{\delta^{'}_{i}\delta^{'}_{i-1}-\delta^{''}_{i}\delta^{''}_{i-1}}{m_i}|\]
  \[=|\frac{\delta^{'}_{i-1}(\delta^{'}_{i}-\delta^{''}_{i})+\delta^{''}_{i}(\delta^{'}_{i-1}-\delta^{''}_{i-1})}{m_i}|\]
  \[\leq|\frac{\delta^{'}_{i-1}(\delta^{'}_{i}-\delta^{''}_{i})}{m_i}|+|\frac{\delta^{''}_{i}(\delta^{'}_{i-1}-\delta^{''}_{i-1})}{m_i}|\]
  \[\leq b\delta^{'}_{i-1}|(\delta^{'}_{i}-\delta^{''}_{i})|+b\delta^{''}_{i}|(\delta^{'}_{i-1}-\delta^{''}_{i-1})|\]
  \[\leq 2b{\|\vec{{\delta}^{'}}-\vec{{\delta}^{''}}\|}_{\infty}\]
  Hence ${\|g(\vec{{\delta}^{'}})-g(\vec{{\delta}^{''}})\|}_{\infty} \leq
  {2b\|\vec{{\delta}^{'}}-\vec{{\delta}^{''}}\|}_{\infty}$.
  \item Part 3:
      $g_{i}(\vec{\delta})=\frac{\delta_{i}\delta_{i+1}}{m_{i+1}}$. The
      proof is similar to part 2. Hence
      ${\|g(\vec{{\delta}^{'}})-g(\vec{{\delta}^{''}})\|}_{\infty} \leq
      {2b\|\vec{{\delta}^{'}}-\vec{{\delta}^{''}}\|}_{\infty}$.
  \item Part 4: $g_{i}(\vec{\delta})=p_k\delta_{i}$. The proof is similar
      to part 1. Hence
      ${\|g(\vec{{\delta}^{'}})-g(\vec{{\delta}^{''}})\|}_{\infty} \leq
      {a\|\vec{{\delta}^{'}}-\vec{{\delta}^{''}}\|}_{\infty}$.
\end{itemize}

Summing the above four parts, we can prove that for $i\neq0$, $1$,
$h_{N}+1$, $h, h_N$ (i.e., in middle lists), there exists a constant $L$
which is independent with $p_{k}$'s and $m_{i}$'s, such that the function
$f$ is Lipschitz-continuous of constant $L(a+b)$ on ${\mathbf{X}}$, that is
\[{\|f(\vec{{\delta}^{'}})-f(\vec{{\delta}^{''}})\|}_{\infty}\leq L(a+b){\|\vec{{\delta}^{'}}-\vec{{\delta}^{''}}\|}_{\infty}\]
where $a =max_{k}p_{k}$ and $b=max_{i}(1/m_{i})$. By summing all the cases
of $i$, we show Lemma 4 holds. \done

\vskip  5px \noindent{\bf Proof of lemma~\ref{app_lemma4}: } The proof of
Lemma \ref{app_lemma4} is simple as $\vec{\delta}(0)$ equals $\vec{H}(0)$
and $f(\vec{\delta}(t))$ is the variation of $\vec{\delta}(t)$ for all $i$.
\done

\vskip  5px \noindent{\bf Proof of Theorem 3: } Let
$\vec{M}(t)=\sum_{s=0}^{t-1}(\vec{H}(s+1)-\vec{H}(s)-f(\vec{H}(s)))$, we
have:
\begin{eqnarray}
\vec{H}(t)
&=&\vec{H}(0)+\sum_{s=0}^{t-1}f(\vec{H}(s))+\vec{M}(t)
\label{eq:theorem2_H}
\end{eqnarray}

Combining Lemma 5 and Equation (\ref{eq:theorem2_H}), we get:
\[\vec{H}(t)-\vec{\delta}(t)=\sum_{s=0}^{t-1}(f(\vec{H}(s))-f(\vec{\delta}(s)))+\vec{M}(t)\]
By using norm, we have, for $t \leq \tau$, ${\|\vec{H}(t)-\vec{\delta}(t)\|}_\infty$ is less than
\[\sum_{s=0}^{\tau-1}{\|(f(\vec{H}(s))-f(\vec{\delta}(s)))\|}_\infty+\sup_{t\leq \tau}{\|\vec{M}(t)\|}_\infty\]
By using Lemma \ref{app_lemma3}, we get
${\|(f(\vec{H}(s))-f(\vec{\delta}(s)))\|}_\infty \leq L(a+b)
{\|\vec{H}(s)-\vec{\delta}(s)\|}_\infty \leq L(a+b)$. Hence, for $t \leq
\tau$, $E[{\|\vec{H}(t)-\vec{\delta}(t)\|}_\infty]$ is less than
\[\sum_{s=0}^{\tau-1}{L(a+b)E[\|(\vec{H}(s)-\vec{\delta}(s))\|}_\infty]+E[\sup_{t\leq \tau}{\|\vec{M}(t)\|}_\infty]\]
By using Lemma \ref{app_lemma2}, $E[{\|\vec{M}(\tau)\|}^2_\infty]\leq
2a^2\tau$.
Besides, we  have

{\small
\begin{eqnarray}
\!\!\!\!&&\!\!\!E[\vec{M}(t+1)|\vec{M}(t),\vec{M}(t-1),...,\vec{M}(0)] \nonumber\\
&=& \!\!\!E[\vec{M}(t)\!+\!\vec{H}(t+1)\!-\!\vec{H}(t)\!-\!f(\vec{H}(t))|\vec{M}(t),\vec{M}(t-1),...,\vec{M}(0)] \nonumber\\
&=& \!\!\!E[\vec{M}(t)|\vec{M}(t),\vec{M}(t-1),...,\vec{M}(0)]\nonumber\\
&+&\!\!\!E[\vec{H}(t+1)-\vec{H}(t)-f(\vec{H}(t))|\vec{M}(t),\vec{M}(t-1),...,\vec{M}(0)]\nonumber\\
&=&\!\!\!E[\vec{M}(t)|\vec{M}(t),\vec{M}(t-1),...,\vec{M}(0)] = \vec{M}(t) \nonumber
\label{eq:app_prooftheorem3_1}
\end{eqnarray}
} So we have $E[\vec{M}(t+1)|\vec{M}(t),\vec{M}(t-1),...,\vec{M}(0)] =
\vec{M}(t), $ which means that $\vec{M}(t)$ is martingale. Thus, we have
\[E[\sup_{t\leq \tau}{\|\vec{M}(\tau)\|}_\infty]\leq E[{\|\vec{M}(\tau)\|}^2_\infty]\leq \sqrt{2a^2\tau} \leq \sqrt{2\tau}(a+b)\]

So far, we can prove  that for $t \leq \tau$,
${E[\|\vec{H}(t)-\vec{\delta}(t)\|}_\infty]$ is less than
$L(a+b)\sum_{s=0}^{\tau-1}{E[\|(\vec{H}(s)-\vec{\delta}(s))\|}_\infty]+\sqrt{2\tau}(a+b)$.
By using Discrete Gronwall inequality in \cite{holte2009discrete}, The above
inequality implies that
$E[\sup_{t\leq\tau}{\|\vec{H}(t)-\vec{\delta}(t)\|}_\infty]$ is less than
$\big(\sqrt{2\tau}(a+b)\big)exp(L(a+b)\tau)$. Now by replacing $\tau$ with
$\frac{1}{L(a+b)}$, we can have $E[\sup_{t\leq
\frac{1}{L(a+b)}}{\|\vec{H}(t)-\vec{\delta}(t)\|}_\infty]$ is less than
$\big(\sqrt{2(a+b)/L}\big)e$. Now we can finally show that when
$p_k\rightarrow 0$ as $n\rightarrow \infty$ ($a=max_{k}p_k \rightarrow 0$)
and $m_i\rightarrow \infty$ ($i=0, 1, ..., h$), then for $\tau
=\frac{1}{L(a+b)} \rightarrow \infty$, $E[\sup_{t\leq
\tau}{\|\vec{H}(t)-\vec{\delta}(t)\|}_\infty] \rightarrow 0$, where
$\vec{\delta}(t)=\sum_{k}x_{k,i}(t)p_k$, $\vec{H}(t)=\sum_{k}X_{k,i}(t)p_k$
and $\vec{H}(0)=\vec{\delta}(0)$. $\done$

%
%

%% file: paper.bbl
\begin{thebibliography}{10}

\bibitem{boudec2010stationary}
J.-Y.~L. Boudec.
\newblock The stationary behaviour of fluid limits of reversible processes is
  concentrated on stationary points.
\newblock {\em arXiv preprint arXiv:1009.5021}, 2010.

\bibitem{bovet2005understanding}
D.~P. Bovet and M.~Cesati.
\newblock {\em {Understanding the Linux Kernel}}.
\newblock O'Reilly Media, Inc., 2005.

\bibitem{breslau1999web}
L.~Breslau, P.~Cao, L.~Fan, G.~Phillips, and S.~Shenker.
\newblock {Web Caching and Zipf-like Distributions: Evidence and Implications}.
\newblock In {\em INFOCOM}, 1999.

\bibitem{choi2012opamp}
J.-H. Choi, S.-M. Kim, C.~Kim, K.-W. Park, and K.~H. Park.
\newblock {OPAMP: Evaluation Framework for Optimal Page Allocation of Hybrid
  Main Memory Architecture}.
\newblock In {\em IEEE, ICPADS'2012}.

\bibitem{dhiman2009pdram}
G.~Dhiman, R.~Ayoub, and T.~Rosing.
\newblock {PDRAM: a Hybrid PRAM and DRAM Main Memory System}.
\newblock In {\em IEEE DAC}, 2009.

\bibitem{transient15}
N.~Gast and B.~Van~Houdt.
\newblock {Transient and Steady-state Regime of A Family of List-based Cache
  Replacement Algorithms}.
\newblock In {\em ACM SIGMETRICS}, 2015.

\bibitem{holland2013flash}
D.~A. Holland, E.~L. Angelino, G.~Wald, and M.~I. Seltzer.
\newblock {Flash Caching on The Storage Client}.
\newblock In {\em USENIX, ATC'2013}.

\bibitem{holte2009discrete}
J.~M. Holte.
\newblock Discrete gronwall lemma and applications.
\newblock In {\em MAA-NCS meeting at the University of North Dakota},
  volume~24, pages 1--7, 2009.

\bibitem{Hu2013Software}
J.~Hu, Q.~Zhuge, C.~J. Xue, W.~C. Tseng, and H.~M. Sha.
\newblock {Software Enabled Wear-leveling for Hybrid PCM Main Memory on
  Embedded Systems}.
\newblock In {\em DATE}, 2013.

\bibitem{flashsystem820}
IBM.
\newblock { IBM FlashSystem 820 and IBM FlashSystem 720}.

\bibitem{Kgil2006FlashCache}
T.~Kgil and T.~Mudge.
\newblock {FlashCache: A NAND Flash Memory File Cache for Low Power Web
  Servers}.
\newblock In {\em CASES}, 2006.

\bibitem{kgil2008improving}
T.~Kgil, D.~Roberts, and T.~Mudge.
\newblock {Improving NAND Flash Based Disk Caches}.
\newblock In {\em ISCA}. IEEE, 2008.

\bibitem{FAST14}
H.~Kim, S.~Seshadri, C.~L. Dickey, and L.~Chiu.
\newblock {Evaluating Phase Change Memory for Enterprise Storage Systems: A
  Study of Caching and Tiering Approaches}.
\newblock In {\em USENIX, FAST'2014}.

\bibitem{lee2009architecting}
B.~C. Lee, E.~Ipek, O.~Mutlu, and D.~Burger.
\newblock {Architecting Phase Change Memory As A Scalable Dram Alternative}.
\newblock {\em ACM SIGARCH Computer Architecture News}, 37(3):2--13, 2009.

\bibitem{lee2014eliminating}
E.~Lee, H.~Kang, H.~Bahn, and K.~Shin.
\newblock {Eliminating Periodic Flush Overhead of File I/O With Non-volatile
  Buffer Cache}.
\newblock In {\em Transactions on Computers}. IEEE, 2014.

\bibitem{ISCA09}
M.~K. Qureshi, V.~Srinivasan, and J.~A. Rivers.
\newblock {Scalable High Performance Main Memory System Using Phase-change
  Memory Technology}.
\newblock {\em ACM SIGARCH Computer Architecture News}, 37(3):24--33, 2009.

\bibitem{raoux2008phase}
S.~Raoux, G.~W. Burr, M.~J. Breitwisch, C.~T. Rettner, Y.-C. Chen, R.~M.
  Shelby, M.~Salinga, D.~Krebs, S.-H. Chen, H.-L. Lung, et~al.
\newblock {Phase-change Random Access Memory: A Scalable Technology}.
\newblock {\em IBM Journal of Research and Development}, 52(4.5):465--479,
  2008.

\bibitem{dramsim2}
P.~Rosenfeld, E.~Cooper-Balis, and B.~Jacob.
\newblock {DRAMSim2: A Cycle Accurate Memory System Simulator}.
\newblock {\em Computer Architecture Letters}, 10(1):16--19, 2011.

\bibitem{xue2011emerging}
C.~J. Xue, Y.~Zhang, Y.~Chen, G.~Sun, J.~J. Yang, and H.~Li.
\newblock {Emerging Non-volatile Memories: Opportunities and Challenges}.
\newblock In {\em CODES+ISSS}. ACM, 2011.

\bibitem{yates1995framework}
R.~D. Yates.
\newblock A framework for uplink power control in cellular radio systems.
\newblock {\em IEEE Journal on selected areas in communications},
  13(7):1341--1347, 1995.

\bibitem{yoon2012row}
H.~Yoon, J.~Meza, R.~Ausavarungnirun, R.~A. Harding, and O.~Mutlu.
\newblock {Row Buffer Locality Aware Caching Policies for Hybrid Memories}.
\newblock In {\em IEEE, ICCD'2012}.

\bibitem{zheng2011fastscale}
W.~Zheng and G.~Zhang.
\newblock {FastScale: Accelerate RAID Scaling by Minimizing Data Migration.}
\newblock In {\em FAST}, 2011.

\bibitem{zipf1929relative}
G.~K. Zipf.
\newblock {Relative Frequency as a Determinant of Phonetic Change}.
\newblock {\em Harvard studies in classical philology}, 40:1--95, 1929.

\end{thebibliography}
